\documentclass[twocolumn]{openjournal}
\usepackage{graphicx}   
\usepackage{amsmath}    
\usepackage{amssymb,amsfonts}    
\usepackage{orcidlink}
\usepackage{verbatim}
\usepackage{hyperref}
\usepackage{xcolor}
\newcommand{\vect}[1]{\boldsymbol{#1}}
 \usepackage{lineno}

\shorttitle{von Zeipel-Kozai-Lidov oscillations in Lambda Ophiuchi}
\shortauthors{I. Waisberg, Y. Klein \& B. Katz}

\begin{document}

\title{von Zeipel-Kozai-Lidov oscillations in nearby bright stars \\ I. Lambda Ophiuchi \thanks{Based on observations collected at the European Southern Observatory, Chile, Program IDs 111.24UP.003, 113.26HF.001}}

\footnotetext[]{Based on observations collected at the European Southern Observatory, Chile, Program IDs 111.24UP.001, 111.264P.001, 113.26HF.001}

\newcommand{\weizmann}{Department of Particle Physics and Astrophysics, Weizmann Institute of Science, Rehovot 76100, Israel}

\email{email: idelwaisberg@gmail.com}

\author{\vspace{-1.2cm}Idel Waisberg\,\orcidlink{0000-0003-0304-743X}$^{1,2}$, Ygal Klein\orcidlink{0009-0004-1914-5821}$^{2}$ \& Boaz Katz\orcidlink{0000-0003-0584-2920}$^{2}$}

\affiliation{$^1$Independent researcher, Lambda Ophiuchi Ltda}
\affiliation{$^2$\weizmann}

\begin{abstract}
The challenge of constraining both the inner and the outer orbits in multiple stars has resulted in a growing abyss between the rich theoretical and the sparse observational studies of von Zeipel-Kozai-Lidov (ZKL) oscillations in stellar systems. Here we solve for the full orbital architecture of the bright intermediate-mass nearby system Lambda Ophiuchi based on astrometric measurements of the outer orbit (period of 129 years) compiled in the Sixth Catalog of Orbits of Visual Binary Stars and new VLTI/GRAVITY interferometric measurements that are used to determine the inner orbit (period of 42 days). The orbits are retrograde and misaligned by either $88.5\pm1.9^{\degr}$ or $113.5\pm1.9^{\degr}$, which in either case results in the inner binary currently undergoing ZKL oscillations. While pure Newtonian point source evolution would have predicted the stars in the inner binary to have merged long ago, in reality the eccentricity oscillations are significantly modulated by general relativistic, tidal and rotational bulge precession. We show that due to the effect of ``slaved'' precession the dynamics can still be solved semi-analytically. We find that the (currently unknown) inclination angles between the stellar spins axes and the inner orbital axis play a very important role in the amplitude of the ZKL oscillations, which is at a minimum $\Delta e = e_{\mathrm{max}} - e_{\mathrm{min}} \simeq 0.15$ and could be as high as $\Delta e \simeq 0.70$. We argue that currently feasible spectroscopic and interferometric observations could allow for a complete and unique dynamical solution for this system. 
\end{abstract}


\section{Introduction}
\label{sec:introduction}

Arguably the most interesting dynamical phenomenon to emerge from the gravitational three-body problem are von Zeipel-Kozai-Lidov (ZKL) oscillations in hierarchical triple systems \citep{Ito19,Kozai62,Lidov62}. If the outer companion to an inner binary is on an orbit with a sufficiently high mutual inclination, it can potentially induce high amplitude eccentricity oscillations in the inner binary that could have a decisive influence on the fate of the system. For instance, the oscillations can significantly enhance tidal dissipation and cause the inner binary to shrink \citep[e.g.][]{Kiseleva98}, merge \citep[e.g.][]{Naoz14} or even directly collide \citep[the latter being more feasible for compact objects such as white dwarfs, e.g.][]{Katz12}. 

The statistical relevance of ZKL oscillations in real stellar systems is still poorly constrained. For example, it was once thought that the fact that nearly all solar-type binaries with short periods ($P \lesssim 3 \text{ days}$) have a hierarchical outer companion was evidence that ZKL oscillations played a major role in creating very close binaries \citep[e.g.][]{Tokovinin06}. However, this may have been a case of correlation rather than causation, as observations of T Tauri stars showed that tight binaries already exist only a few Myrs after star formation \citep[e.g.][and references therein]{Moe18} and statistics of the mutual orientation between the inner and outer orbits in hierarchical triples containing a low mass (solar mass and below) close eclipsing binary have show that the majority of such systems are nearly coplanar or only mildly misaligned \citep[e.g.][]{Borkovits16}, which is not consistent with ZKL oscillations having caused the migration. The interpretation is that in order to tighten the binary orbit to short periods a massive circumbinary disk is needed, which in turn is more likely to fragment into additional outer component(s) \citep[e.g.][]{Tokovinin20}. 

For stars of intermediate-mass and above, whether there exists a significant channel of forming close binaries through ZKL oscillations is still an open question; despite great strides in constraining the multiplicity properties of massive stars \citep[e.g.][]{Offner23}, the properties of their orbital architecture (in particular mutual orientations) are still essentially unconstrained. Nonetheless, there are hints that triple systems with more massive primaries tend to be more misaligned compared to lower mass systems \citep[e.g.][]{Gardner22}, which suggests that ZKL oscillations might play a more important role in the case of more massive stars.


ZKL oscillations happen on a timescale of the order $\frac{P_{\mathrm{out}}^2}{P_{\mathrm{in}}}$, where $P_{\mathrm{out}}$ and $P_{\mathrm{in}}$ are the outer and inner orbital periods respectively. For the vast majority of systems which have large hierarchies, the resulting timescales are so long that measuring actual changes in the orbital parameters due to ZKL is virtually impossible. Fortunately, ZKL oscillations can be rather straightforwardly calculated as long as the component masses and orbital parameters of both the inner and outer orbits are sufficiently well constrained. In practice, doing so for real systems is quite challenging as the outer orbits usually have periods of the order of decades to centuries (and therefore require dedicated observational campaigns over long timescales) and the inner orbits usually have periods of the order of months or days (and therefore are challenging to spatially resolve). As a result, the number of stellar systems for which triple dynamics has been constrained is extremely scarce. While a few examples do exist \citep[such as compact triple systems containing an eclipsing binary, e.g.][]{Borkovits22}, we are not aware of \textit{any} stellar system with a large mutual inclination $i_{\mathrm{mut}}\sim90^{\degr}$ that has been proven to be undergoing high amplitude ZKL oscillations\footnote{\textit{Algol}=$\beta$ Per is well known for being a triple system with a mutual inclination very close to $90^{\degr}$ \citep{Baron2012}. However, tidal precession in the very close inner binary completely damps ZKL oscillations in this case.}. 

Nearby and bright visual binaries that turn out to be hierarchical multiple systems are particularly promising targets for dynamical studies. If their outer orbits have a period of about two centuries or less, they have a good chance of having been well covered by historical astrometric observations starting around the early 1800s. Furthermore, being nearby and bright their inner orbits can be easily resolved with optical/near-infrared interferometry even for orbital periods as short as a few days\footnote{We note that spectroscopic orbits do not suffice for dynamical studies because they are insensitive to the longitude of the ascending node $\Omega$, which is necessary to measure the mutual inclination.}. Intermediate-mass multiple systems (that is, those whose primary has a mass $1.5 M_{\odot} \lesssim M \lesssim 8 M_{\odot}$) are particularly interesting for dynamical studies because they make up for the majority of the progenitors of white dwarfs in multiple systems, whose properties are currently only poorly constrained but may play a key role in the supernova type Ia progenitor problem \citep[e.g.][]{Ruiter25}.

In real multiple systems, it is possible and perhaps likely that ZKL oscillations are modulated by precession of the argument of pericenter in the inner binary; relevant contributions include general relativistic precession, tidal precession and rotational bulge precession. This implies that to uniquely solve for the dynamics in a given system one may also often need good constraints on stellar properties such as radii, apsidal constants, rotational velocities and orientation of the stellar spin axes. Just the same, measurements of the orbital parameters for the outer and inner orbits can already provide important constraints on the dynamics of the system. 

Lambda Ophiuchi ($\lambda$ Oph, HIP 80883, HD 148857, HR 6149, 10 Oph) is a nearby and bright (V=3.9) intermediate-mass visual binary (WDS J16309+0159AB). The outer orbit A+B with period $P_{\mathrm{out}} \approx 129 \text{ yr}$ and semi-major axis $a_{\mathrm{out}}\approx0.9" \leftrightarrow 46 \text{ au}$ has been covered by historical observations dating back to 1825 and collected in the Sixth Catalog of Orbits of Visual Binary Stars\footnote{\href{http://www.astro.gsu.edu/wds/orb6.html}{http://www.astro.gsu.edu/wds/orb6.html}} (hereafter ORB6). The visual primary (of IAU-approved proper name \textit{Marfik}) had been suspected to itself be a close binary \citep{Abt80,Heintz93} and we have reported the discovery of its companion at a projected separation of $6 \text{ mas} \leftrightarrow 0.3 \text{ au}$ in a near-infrared VLTI/GRAVITY interferometric observation, while the visual secondary was confirmed to be a single star \citep{Waisberg23}. From isochrone fitting based on photometry and the interferometric flux ratio we estimated masses $M_{Aa} \simeq 2.4 M_{\odot}$, $M_{Ab} \simeq 1.6 M_{\odot}$ and $M_B \simeq 1.8 M_{\odot}$ and a system age of 400 Myr. Interestingly, $\lambda$ Oph also has a very wide common parallax and proper motion companion (C) with mass $M_C \simeq 0.6 M_{\odot}$ at a projected separation $\rho = 120" \leftrightarrow 6400 \text{ au}$. We adopt the latter's precise distance $d=51.3\pm0.1 \text{ pc}$ in Gaia DR3 \citep{Gaia23} for the system since $\lambda$ Oph is currently unsolved in Gaia and its Hipparcos distance is less precise \citep[$d=53.1\pm1.6\text{ pc}$; ][]{vanLeeuwen07}. The projected rotational velocity of the visual primary A is $v \sin i \simeq 140 \text{ km}\text{ s}^{-1}$ \citep{Abt95,Royer02}. 

With the goal of solving for the triple dynamics in $\lambda$ Oph we have obtained a further five new VLTI/GRAVITY observations of the inner binary. This paper is organized as follows. In Section \ref{sec:outer_orbit}, we find an updated solution for the outer binary A+B based on ORB6 measurements. In Section \ref{sec:inner_orbit}, we use the VLTI/GRAVITY measurements to solve for the orbit of the inner binary Aa+Ab. In Section \ref{sec:dynamics}, we solve for the dynamics of the triple system including the ZKL oscillations and the relevant precession terms in the inner binary; in particular, we show that ``slaved'' precession in the inner binary allows for fast semi-analytical solutions even in the presence of misaligned stellar spin axes. In Section \ref{sec:discussion} we discuss the results in the context of the formation and evolution of $\lambda$ Oph, the possible role of the wide companion C, the prospects for obtaining a unique dynamical solution for this system and why the combination of historical outer orbits with interferometric inner orbits should remain the main method for solving the dynamics in intermediate-mass multiple systems. A brief conclusion can be found in Section \ref{sec:conclusion}. 

\section{The outer orbit}
\label{sec:outer_orbit}

The visual A+B orbit with $P_{\mathrm{out}} = 129 \text{ yr}$ has been very well covered by several hundred astrometric observations dating back to 1825. Such observations are collected in ORB6 and were kindly provided to us by Dr. Rachel Matson of the US Naval Observatory. The current orbital solution quoted in ORB6 is that from \cite{Izmailov19}. In order to include the most recent ORB6 measurements up to 2024, we performed our own orbital fit following the same procedure detailed in \cite{Waisberg25} to treat the measurements in ORB6 (which are very heterogeneous and often do not have reported uncertainties). We excluded observations which did not contain both a separation and a position angle measurement and divided the remaining observations in a group of low-resolution observations (mostly micrometry) and high-resolution observations (mostly speckle interferometry, with some adaptive optics and a few older visual interferometry observations). An orbital fit was then performed for each group (all epochs having equal weight) in order to estimate the average astrometric error by requiring the average residual squared to be equal to the square of the astrometric error. At this point, blatant outliers (defined as the residual in either coordinate being larger than four times the estimated error) were excluded and the orbital fit was then repeated and the resulting errors re-estimated. The orbital fit was then performed by combining the two groups with their appropriate errors. The final orbital fit contained 691 low-resolution observations with an estimated error of 88.5 mas and 101 high-resolution observations with an estimated error of 13.3 mas. 

\begin{figure*}[]
 \includegraphics[width=2\columnwidth]{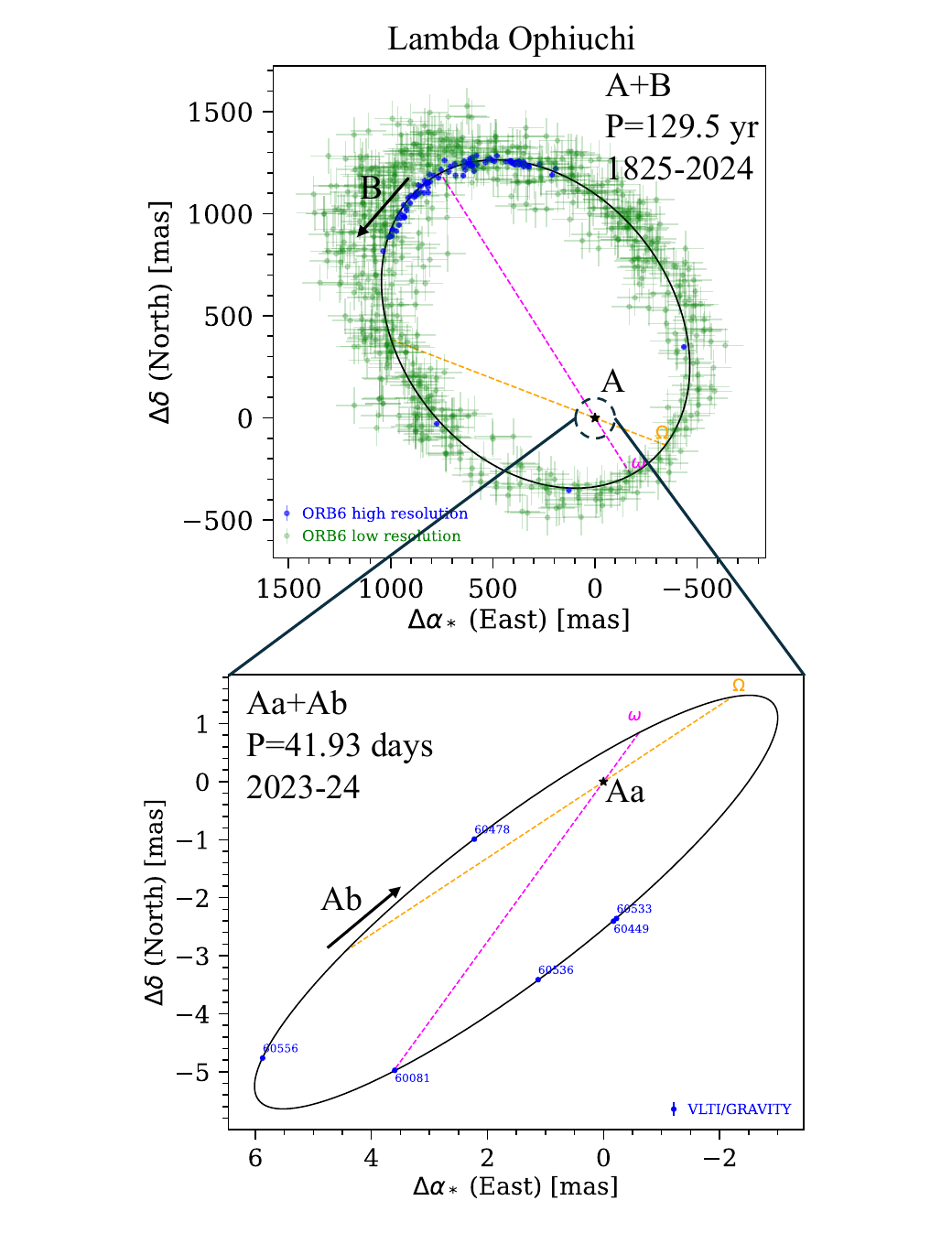}
 \caption{ \label{fig:orbital_fits} Astrometric data (colored) and best fit orbital solution (solid black) for Lambda Ophiuchi A + B (top) and Aa + Ab (bottom). The arrows show the direction of revolution of the secondaries in the sky plane. The line of apsides ($\omega$) and the line of nodes ($\Omega$) are shown in magenta and orange respectively. The numbers next to each point in the lower plot show the observation date (MJD).}
\end{figure*}

Figure \ref{fig:orbital_fits} (top) shows the ORB6 data (colored) together with the best fit orbital solution (black), while Table \ref{table:orbital_fits} reports the best fit parameters and their uncertainties. The latter (defined as the 2.3\% and 97.7\% percentiles in the distribution, corresponding to 2$\sigma$ limits for a Gaussian distribution) were estimated by generating and fitting $2 \times 10^3$ random resamples of the astrometric data according to their estimated errors to get the distribution of the best fit parameters, which is shown in Figure \ref{fig:corner_A+B} in Appendix \ref{app:orbital_params_distributions}. In particular, we note a strong correlation between the argument of pericenter $\omega$ and the longitude of the ascending node $\Omega$ (of the form $\omega +\Omega \approx 70^{\degr}$), which results in larger relative uncertainties compared to the other parameters. Note that without radial velocities there is a perfect degeneracy between two solutions with $(\Omega, \omega)$ and $(\Omega + 180^{\degr}, \omega + 180^{\degr})$. We report the solution with $\Omega < 180^{\degr}$ as is standard practice.

We found that A+B has an orbital semi-major axis $a = 898.4\pm6.0 \text{ mas} \leftrightarrow 46.1\pm0.3 \text{ au}$ and a period $P=129.51\pm0.40 \text{ yrs}$, which imply a dynamical mass $M_{A+B}=5.85\pm0.13 M_{\odot}$, and a relatively high eccentricity $e=0.623\pm0.005$, which imply a periastron distance $a_p = 17.40\pm0.25 \text{ au}$. 

\begin{table}
\centering
\caption{\label{table:orbital_fits} Best fit Keplerian parameters for the outer (A+B) and inner (A$a$+A$b$) orbits of Lambda Ophiuchi.}
\begin{tabular}{ccc}
\hline \hline
& \shortstack{A+B} & \shortstack{Aa+Ab} 
\\ [0.3cm]

\shortstack{a\\(mas)\\(au)} 
& \shortstack{$898.4\pm6.0$\\$46.1\pm0.3$} & \shortstack{$7.17\pm0.01$\\$0.368\pm0.006$} 
\\ [0.3cm]

e 
& $0.623\pm0.005$ & $0.71\pm0.01$ 
\\ [0.3cm]

\shortstack{$i$\\(deg)} 
& $21.9\pm1.6$ & $101.6\pm0.2$ 
\\ [0.3cm]

\shortstack{$\Omega$\\(deg)} 
& $68.9\pm5.3$ & $123.3\pm0.4$ 
\\ [0.3cm]

\shortstack{$\omega$\\(deg)} 
& $141.3\pm5.3$ & $118.1\pm1.0$ 
\\ [0.3cm]

P
& \shortstack{$129.51\pm0.40$\\yrs} & \shortstack{$41.93\pm0.01$\\days}
\\ [0.3cm]

$T_p$ & $1939.40\pm0.20$ & \shortstack{$60102.25\pm0.18$\\MJD} 
\\ [0.3cm]

\hline \\ [0.1cm]

\shortstack{$a_p$\\(au)} 
& $17.40\pm0.25$ & $0.106\pm0.003$ 
\\ [0.3cm]

\shortstack{$M_{\mathrm{dyn}}$\\($M_{\odot}$)} 
& $5.85\pm0.13$ & $3.78\pm0.17$ 
\\ [0.3cm]

\hline
\end{tabular}
\tablenotetext{0}{Notes:}
\tablenotetext{0}{The uncertainties correspond to the average of the 2.3\% and 97.7\% percentiles of the parameter distributions.}
\tablenotetext{0}{For the physical semi-major axes and dynamical masses, the adopted distance is $d=51.3\pm0.1 \text{ pc}$.}
\tablenotetext{0}{Alternative solutions with $(\Omega, \omega) \leftrightarrow (\Omega + 180^{\degr} , \omega + 180^{\degr})$ are possible in both cases.}
\end{table}

\section{The inner orbit}
\label{sec:inner_orbit}

\subsection{VLTI/GRAVITY observations}

In addition to the discovery epoch in 2023 reported in \cite{Waisberg23}, we have obtained a further five VLTI/GRAVITY \citep{GRAVITY17} observations of the visual primary $\lambda$ Oph A throughout 2024 in order to measure the orbital parameters of Aa+Ab. All the observations were performed using the four 1.8-m Auxiliary Telescopes (ATs) in single-field mode, wherein half of the photons were used to fringe track at low spectral resolution (R=22) and the other half were integrated in 10s exposures in high spectral resolution (R=4,000). The angular resolution corresponding to the largest projected baseline varied from 2.3 to 4.8 mas between the epochs. In each epoch two files were obtained, each containing 16 exposures. Table \ref{table:observations_gravity} in Appendix \ref{app:gravity_observations} provides details of the observations, including the largest projected baselines $B_{\mathrm{proj,max}}$ and corresponding angular resolutions $\theta_{\mathrm{max}}$. For completeness, the discovery epoch in 2023 is also included in Table \ref{table:observations_gravity}. In all the epochs the $\lambda$ Oph A observations were followed by observations of the G8III star HD 148287 (angular diameter 0.73 mas), which served as the interferometric calibrator. The data were reduced with the ESO GRAVITY pipeline v.1.6.6 \citep{Lapeyrere14}. 

\subsection{Astrometric measurements}

The VLTI/GRAVITY interferometric data were fit with the binary model detailed in \cite{Waisberg23b} in order to derive the astrometry for the Aa+Ab orbit. The model parameters are the K band flux ratio between Ab and Aa ($\frac{f_{Ab}}{f_{Aa}}$) and the projected separation of Ab relative to Aa in the East and North directions $(\Delta \alpha_*, \Delta \delta)_{\mathrm{Ab,Aa}}$. The angular diameters of the stars were fixed to $\theta_{Aa} = 0.43 \text{ mas}$ and $\theta_{Ab} = 0.27 \text{ mas}$ based on the radii estimated from isochrone fitting ($R_{Aa} = 2.5 R_{\odot}$, $R_{Ab} = 1.6 R_{\odot}$) in \cite{Waisberg23} and have a negligible effect on the binary model fit since they are well below the interferometric resolution. 

\begin{figure}[]
\includegraphics[width=\columnwidth]{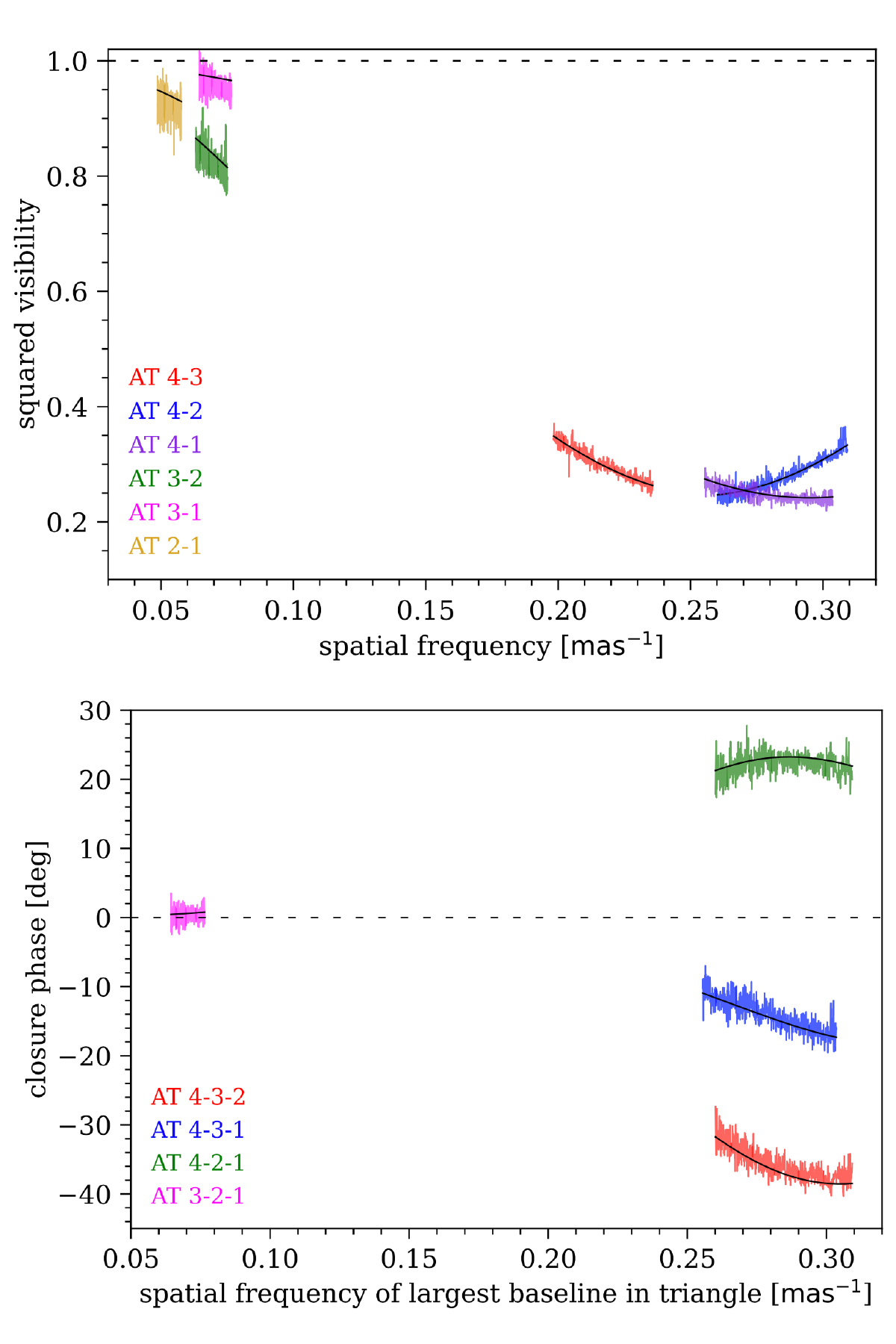}
\caption{\label{fig:gravity_fit_binary} VLTI/GRAVITY data (colored) for $\lambda$ Oph Aa+Ab and best fit binary model (solid black) for epoch 2024-05-19. The upper panel shows the squared visibilities for the six baselines and the lower panel shows the closure phases for the four triangles. The dashed lines show the expected values for a single unresolved star.}
\end{figure}

The binary model fitting results for each epoch are reported in Table \ref{table:observations_gravity}. The K band flux ratio is $\frac{f_{Ab}}{f_{Aa}} = 33.0 \pm 0.5 \%$ based on the mean and the standard deviation for all the epochs. The formal astrometric errors for each epoch are on the order of a few microarcseconds and are expected to be underestimated due to correlations between spectral channels; more realistic errors are estimated below based on the orbital fit. Figure \ref{fig:gravity_fit_binary} shows the interferometric data and best-fit binary model for the first epoch in 2024. Corresponding figures for the other epochs can be found in Appendix \ref{app:gravity_observations}. 

\subsection{Orbital fit}

Armed with the six VLTI/GRAVITY astrometric measurements, we proceeded to find the best-fit orbital solution for Aa+Ab. In order to find the global minimum, we ran several fits over a three-dimensional grid in eccentricity $0 < e < 1$ with steps of 0.05, period $20 < P < 60 \text{ days}$ in steps of 2 days and time of pericenter (MJD) $60081.210 - P/2 < T_p < 60081.210 + P/2$ in steps of 1 day. The average astrometric error, estimated by requiring the average residual squared to be equal to the square of the astrometric error, is 0.016 mas in each direction.

Figure \ref{fig:orbital_fits} (bottom) shows the measured astrometric positions (also labeled by their MJD) together with the best fit orbit. The best fit parameters and their uncertainties are reported in Table \ref{table:orbital_fits}. The uncertainties (defined as the 2.3\% and 97.7\% percentiles) were estimated using the same procedure as detailed above for the outer orbit and their full distributions are plotted in Figure \ref{fig:corner_Aa+Ab} in Appendix \ref{app:orbital_params_distributions}. As in the case of the outer orbit, the lack of radial velocities lead to a perfect degeneracy between $(\Omega, \omega)$ and $(\Omega + 180^{\degr}, \omega + 180^{\degr})$, so we report the solution with $\Omega < 180^{\degr}$ as is standard practice. 

We find that the orbital parameters of Aa+Ab are extremely well constrained without any strong correlations despite there being only six measurement epochs. This is due to a decent orbital phase coverage combined with the extremely high precision of the interferometric observations. In particular, we note the two epochs with almost coincident positions (MJD=60449 and 60533) that happened to have very similar orbital phases but are separated in time by two orbital periods. 

Aa+Ab has an orbital semi-major axis $a = 7.17\pm0.01 \text{ mas} \leftrightarrow 0.368\pm0.006 \text{ au}$ and a period $P=41.93\pm0.01 \text{ days}$, which imply a dynamical mass $M_{Aa+Ab}=3.78\pm0.17 M_{\odot}$, and an eccentricity $e=0.71\pm0.01$, which imply a periastron distance $a_p = 0.106\pm0.003 \text{ au}$. 

In \cite{Waisberg23} we used isochrones to estimate a system age of 400 Myr and component masses $M_{Aa} \simeq 2.4 M_{\odot}$, $M_{Ab} \simeq 1.6 M_{\odot}$ and $M_B \simeq 1.8 M_{\odot}$. These masses yield $M_{Aa+Ab} \simeq 4.0 M_{\odot}$ and $M_{A+B} \simeq 5.8 M_{\odot}$ and are consistent with the dynamical masses we measured for both the inner and outer orbits, so we adopt them in this paper as well. 

Finally, we can estimate the effect of the inner orbit on the astrometric measurements of the outer orbit. The semi-major axis of the inner orbit's photocenter around the inner orbit's center of mass (which is what truly traces a Keplerian orbit around component B) is 

\begin{align}
a_{\mathrm{phot}} = a_{Aa} \frac{1-f/q}{1+f} = \frac{q}{1+q} a_{\mathrm{in}} \frac{1-f/q}{1+f}
\end{align}

\noindent where $q=\frac{M_{\mathrm{Ab}}}{M_{\mathrm{Aa}}} \simeq 0.67$ is the mass ratio and $f=\frac{f_{\mathrm{Ab}}}{f_{\mathrm{Aa}}}$ is the flux ratio in a given band. In the K band we have $f=0.33$ so that $a_{\mathrm{phot}} \simeq 1.1 \text{ mas}$. In the V band (in which most of the outer orbit's astrometric measurements were made) we estimate $f=0.19$ based on the isochrone solution in \cite{Waisberg23} so that $a_{\mathrm{phot}} \simeq 1.7 \text{ mas}$. The maximum separation between the photocenter and the center of mass is thus $a_{\mathrm{phot}} (1+e) \simeq 2.9 \text{ mas}$, which is only about $20\%$ of the estimated astrometric error even for the high-resolution measurements in ORB6. We therefore conclude that the photocenter motion of A due to the Aa+Ab orbit only has a very minor contribution to the astrometric errors in the outer orbit measurements and therefore there is no benefit to include it in the outer orbit fit.

\section{Triple dynamics}
\label{sec:dynamics}

\subsection{Mutual inclination}

The mutual inclination $i_{\mathrm{mut}}$ is a key parameter in the dynamics of hierarchical triple systems in the sense that a large $i_{\mathrm{mut}}$ is required (although not sufficient) for large amplitude ZKL oscillations in the inner binary. In the case of Lambda Ophiuchi, our measured mutual inclination 

\begin{align}
\begin{split}
&i_{\mathrm{mut}} = \arccos ( \hat{j}_{\mathrm{in}} \cdot \hat{j}_{\mathrm{out}}) = \\&\arccos (\cos i_{\mathrm{in}} \cos i_{\mathrm{out}} + \sin i_\mathrm{in} \sin i_{\mathrm{out}} \cos (\Omega_\mathrm{in} - \Omega_{\mathrm{out}}) )
\end{split}
\end{align}

\noindent has two possible values due to the $(\Omega, \omega) \leftrightarrow (\Omega + 180^{\degr}, \omega + 180^{\degr})$ degeneracy in our orbital solutions ($\hat{j}$ is the unit vector in the direction of the orbital angular momentum axis). Namely, for

\begin{align}
\begin{split}
&(\Omega_{\mathrm{in}}, \Omega_{\mathrm{out}}) = 
\begin{cases}
(123.3\pm0.4^{\degr},68.9\pm5.3^{\degr}) \\
\mathrm{or} \\
(303.3\pm0.4^{\degr},248.9\pm5.3^{\degr})
\end{cases}
\\ &\Rightarrow i_{\mathrm{mutual}} = 88.5\pm1.9^{\degr}
\end{split}
\end{align}

\noindent while for 

\begin{align}
\begin{split}
&(\Omega_{\mathrm{in}}, \Omega_{\mathrm{out}}) = 
\begin{cases}
(123.3\pm0.4^{\degr},248.9\pm5.3^{\degr}) \\
\mathrm{or} \\
(303.3\pm0.4^{\degr},68.9\pm5.3^{\degr})
\end{cases}
\\ &\Rightarrow i_{\mathrm{mutual}} = 113.5\pm1.9^{\degr}
\end{split}
\end{align}

\noindent so that both possible mutual inclinations are large and favorable for possibly interesting dynamical behavior. 

It is also interesting to note that the inner and outer orbits are retrograde (i.e. $i_{\mathrm{in}} < 90^{\degr}$ and $i_{\mathrm{out}} > 90^{\degr}$). 

\subsection{Timescales and relevant effects}

In ZKL oscillations the torque from the outer component causes the argument of pericenter, eccentricity and mutual inclination of the inner binary to oscillate while the Hamiltonian and total angular momentum must be kept constant. The ZKL oscillations can be suppressed if other contributions to the apsidal precession of the inner binary have a comparable or shorter timescale since this causes the torque from the outer companion to average out. In the case of Lambda Ophiuchi, the relevant contributions are due to General Relativity (GR), tidal bulge and rotational bulge. Expressions for the timescales can be found for e.g. in \cite{Fabrycky07}. 

The ZKL timescale is given by 

\begin{align}
\tau_{\mathrm{ZKL}} = \frac{8}{15 \pi} \frac{m_{Aa}+m_{Ab}+m_B}{m_B} \frac{P_{\mathrm{out}}^2}{P_{\mathrm{in}}} (1-e_{\mathrm{out}}^2)^{3/2} \approx 130 \text{ kyr}
\end{align}

GR precession occurs with a timescale 

\begin{align}
&\dot{\omega}_{\mathrm{GR}} = \frac{3 (G (m_{Aa} + m_{Ab}))^{3/2}}{a_{\mathrm{in}}^{5/2} c^2 (1-e_{\mathrm{in}}^2)} \\
&\tau_{\mathrm{GR}} = \frac{2 \pi}{\dot{\omega}_{\mathrm{GR}}} \approx 175 \text{ kyr}
\end{align}

\noindent where $G$ is Newton's constant and $c$ is the speed of light. 

Tidal precession occurs at a rate 

\begin{align}
\begin{split}
&\dot{\omega}_{\mathrm{tide}} = \left (\frac{G (m_{Aa}+m_{Ab})}{a_{\mathrm{in}}^3} \right )^{1/2} 15 \frac{1+\frac{3}{2}e_{\mathrm{in}}^2+\frac{1}{8}e_{\mathrm{in}}^4}{(1-e_{\mathrm{in}}^2)^5} \\ &\left ( k_1 \frac{m_{Ab}}{m_{Aa}} \left ( \frac{R_{Aa}}{a_{\mathrm{in}}} \right )^5 + k_2 \frac{m_{Aa}}{m_{Ab}} \left ( \frac{R_{Ab}}{a_{\mathrm{in}}} \right )^5 \right ) \\ 
\end{split} \\
&\tau_{\mathrm{tide}} = \frac{2 \pi}{\dot{\omega}_{\mathrm{tide}}} \approx 1650 \text{ kyr} 
\end{align}

\noindent where $R_{Aa} = 2.5 R_{\odot}$ and $R_{Ab} = 1.5 R_{\odot}$ are the stellar radii and $k_{Aa} = 0.003$ and $k_{Ab} = 0.004$ are the apsidal constants. Our adopted radii come from the best-fit isochrone solution in \cite{Waisberg23}, while the apsidal constants were estimated based on the theoretical model grids in \cite{Claret23} with an age of 400 Myr, Z=0.0134 (solar metallicity) and masses of $2.50 M_{\odot}$ and $1.60 M_{\odot}$ for the primary and secondary, respectively. Comparison between theoretical and empirical apsidal constants based on eccentric eclipsing binaries have consistently found excellent agreement \citep[e.g.][]{Claret21}. Furthermore, using equation (4) in \citep{Claret23} we found that the correction to the apsidal constant due to rotation is quite minor in our case (e.g. less than $10\%$ smaller for $v \sim 140 \text{ km}\text{ s}^{-1}$ in the case of Aa).

Finally, the precession due to the rotational bulge of the stars depends on the rotational velocities as well as on the unknown angles between the orbital and spin axes (for large misalignment angles, it may even be negative). In the case of alignment we have that 

\begin{align}
\begin{split}
&\dot{\omega}_{\mathrm{rotate}} = \left (\frac{a_{\mathrm{in}}^3}{G (m_{Aa}+m_{Ab})} \right )^{1/2} \frac{1}{(1-e_{\mathrm{in}}^2)^2} \\ &\left ( k_{\mathrm{Aa}} \left ( 1 + \frac{m_{Ab}}{m_{Aa}} \right ) \left ( \frac{R_{Aa}}{a_{\mathrm{in}}} \right )^5 \Omega_{\mathrm{rotate,Aa}}^2 \right. + \\ & \left. k_{\mathrm{Ab}} \left ( 1 + \frac{m_{Ab}}{m_{Aa}} \right ) \frac{m_{Aa}}{m_{Ab}} \left ( \frac{R_{Ab}}{a_{\mathrm{in}}} \right )^5 \Omega_{\mathrm{rotate,Ab}}^2 \right ) \\ 
\end{split} \\
&\tau_{\mathrm{rotate}} = \frac{2 \pi}{\dot{\omega}_{\mathrm{rotate}}} \sim 60 \text{ kyr} 
\end{align}

\noindent where $\Omega_{\mathrm{rotate}} = \frac{v_{\mathrm{rotate}}}{R}$ and $v_{\mathrm{rotate}} \sim v \sin i = 140 \text{ km}\text{ s}^{-1}$. 

To summarize, in the case of $\lambda$ Oph we have that $\tau_{\mathrm{rotate}} \lesssim \tau_{\mathrm{ZKL}} \sim \tau_{\mathrm{GR}} \ll \tau_{\mathrm{tide}}$. It is curious that $\tau_{\mathrm{rotate}}$, $\tau_{\mathrm{ZKL}}$ and $\tau_{\mathrm{GR}}$ are comparable in this system. Furthermore, if the eccentricity can increase significantly then $\tau_{\mathrm{tide}}$ could also shorten significantly compared to the other timescales since it has the strongest dependence on $e_{\mathrm{in}}$. Therefore, all of these terms should be taken into account to study the real dynamics of the system.

\subsection{Analytical solution for the triple dynamics}

The range of eccentricities and inclinations spanned by the oscillations can actually be found by using the conservation of energy and total angular momentum without the need to integrate either the orbit or the orbital parameters. Given the high hierarchy of $\lambda$ Oph, the double-averaged (i.e. averaged over both outer and inner orbits) quadrupolar approximation to the perturbing potential of the outer body is a very good one. For instance, the relative strength of the octupole term relative to the quadrupole term of the perturbing potential, $\epsilon_{\mathrm{oct}}$, is 

\begin{align}
\epsilon_{\mathrm{oct}} = \frac{M_{Aa}-M_{Ab}}{M_{Aa}+M_{Ab}} \frac{a_{\mathrm{in}}}{a_{\mathrm{out}}} \frac{e_{\mathrm{out}}}{1-e_{\mathrm{out}}^2} \simeq 0.0016
\end{align} 

\noindent \citep[e.g.][]{Naoz16}. Given that $\epsilon_{\mathrm{oct}} \ll 1$, it can be safely neglected over the next several Kozai oscillations of the system.

The double-averaged quadrupolar term of the perturbing potential is given by

\begin{align}
\label{eq:H_quad}
\begin{split}
&H_{\mathrm{quad}} = - \frac{G m_{Aa} m_{Ab} m_B}{m_{Aa} + m_{Ab}} \frac{a_{\mathrm{in}^2}}{8 a_{\mathrm{out}^3} (1-e_{\mathrm{out}}^2)^{3/2}} \\ &\times (2 + 3 e_{\mathrm{in}}^2 - 3 \sin^2 i_{\mathrm{mut}} + 3 e_{\mathrm{in}}^2 \sin^2 i_{\mathrm{mut}} - 15 e_{\mathrm{in}}^2 \sin^2 \bar{\omega_{\mathrm{in}}} \sin^2 i_{\mathrm{mut}} )
\end{split}
\end{align}

\noindent where $i_{\mathrm{mut}}$ is the relative inclination between the inner and outer orbits ($\cos i_{\mathrm{mut}} = \hat{j}_{\mathrm{in}} \cdot \hat{j}_{\mathrm{out}}$) and $\bar{\omega}_{\mathrm{in}}$ is the argument of pericenter of the inner orbit \textit{in a coordinate system oriented with the outer orbit} i.e. with $\hat{z} = \hat{j}_{\mathrm{out}}$ \citep[e.g.][]{Fabrycky07}. The last point is important because the standard observational argument of pericenter $\omega$ (e.g. that we report in Table \ref{table:orbital_fits}) is defined in a coordinate system in which the $\hat{z}$ axis is aligned with the line of sight towards the target. It is useful to rewrite Eq \ref{eq:H_quad} in ``covariant'' form by noting that 

\begin{align}
\sin i_{\mathrm{rel}} \sin \bar{\omega}_{\mathrm{in}} = \hat{e}_{\mathrm{in}} \cdot \hat{z} = \hat{e}_{\mathrm{in}} \cdot \hat{j}_{\mathrm{out}}
\end{align}

\noindent so that 

\begin{align}
\label{eq:H_quad_covariant}
\begin{split}
&H_{\mathrm{quad}} = - \frac{G m_{Aa} m_{Ab} m_B}{m_{Aa} + m_{Ab}} \frac{a_{\mathrm{in}^2}}{8 a_{\mathrm{out}^3} (1-e_{\mathrm{out}}^2)^{3/2}} \\ &\times (2 + 3 e_{\mathrm{in}}^2 - 3 \sin^2 i_{\mathrm{mut}} + 3 e_{\mathrm{in}}^2 \sin^2 i_{\mathrm{mut}} - 15 e_{\mathrm{in}}^2 ( \hat{e_{\mathrm{in}}} \cdot \hat{j_{\mathrm{out}}})^2 ) 
\end{split}
\end{align}

Eq. \ref{eq:H_quad_covariant} can now be used with the standard orbital parameters and we recall that in any given coordinate system

\begin{align}
\hat{j} =
\begin{pmatrix}
\sin i \sin \Omega \\
- \sin i \cos \Omega \\
\cos i 
\end{pmatrix} 
; 
\hat{e} = 
\begin{pmatrix}
-\sin \omega \cos i \sin \Omega + \cos \omega \cos \Omega \\
\sin \omega \cos \Omega \cos i + \sin \Omega \cos \omega  \\
\sin i \sin \omega 
\end{pmatrix}
\end{align} 

If tidal dissipation is negligible, the Hamiltonian should be conserved. Another conserved quantity is the total angular momentum vector. If tidal synchronization is negligible, we have that 

\begin{align}
\label{eq:J_tot}
\begin{split}
&J_{\mathrm{total}}^2 = | \vect{J}_{\mathrm{in}} + \vect{J}_{\mathrm{out}} |^2 = \\ &J_{\mathrm{in}}^2 + 2 J_{\mathrm{in}} J_{\mathrm{out}} \cos i_{\mathrm{mut}} + J_{\mathrm{out}}^2 = \text{ constant} 
\end{split} 
\end{align}

\noindent where $J_{\mathrm{in}}$ and $J_{\mathrm{out}}$ are the magnitudes of the orbital angular momentum of the inner and outer orbits: 

\begin{align}
\begin{split}
&J_{\mathrm{in}} = \frac{m_{Aa} m_{Ab}}{m_{Aa}+m_{Ab}} \sqrt{G (m_{Aa} + m_{Ab}) a_{\mathrm{in}} (1-e_{\mathrm{in}}^2)} \\ = &\mu_{\mathrm{in}} \sqrt{G m_{\mathrm{in}} a_{\mathrm{in}} (1-e_{\mathrm{in}}^2)} 
\end{split}
\end{align} 

\begin{align}
\begin{split}
& J_{\mathrm{out}} = \frac{(m_{Aa}+m_{Ab}) m_B}{(m_{Aa} + m_{Ab}) + m_B } \\ &\times \sqrt{G (m_{Aa} + m_{Ab} + m_B) a_{\mathrm{out}} (1-e_{\mathrm{out}}^2)} \\ &= \mu_{\mathrm{out}} \sqrt{G m_{\mathrm{out}} a_{\mathrm{out}} (1-e_{\mathrm{out}}^2)}
\end{split}
\end{align}

\noindent where $\mu$ is the reduced mass. By noting that the magnitude $J_{\mathrm{out}}$ is constant (since both $a_{\mathrm{out}}$ and $e_{\mathrm{out}}$ are constant within the quadrupolar approximation) and that $a_{\mathrm{in}}$ is also constant (in the absence of tidal dissipation), Eq. \ref{eq:J_tot} further simplifies to 

\begin{align}
\label{eq:J_tot_conservation}
\epsilon (1-e_{\mathrm{in}}^2) + \sqrt{1-e_{\mathrm{in}}^2} \cos i_{\mathrm{mut}} = \text{ constant} \\ 
\epsilon = \frac{\mu_{\mathrm{in}}\sqrt{G m_{\mathrm{in}} a_{\mathrm{in}}}}{2 J_{\mathrm{out}}}
\end{align}

In passing, we note that the canonical Kozai-Lidov condition 

\begin{align} 
\label{eq:Kozai_condition}
\sqrt{1-e_{\mathrm{in}}^2} \cos i_{\mathrm{mut}} = \text{ constant}
\end{align}

\noindent only strictly applies in the test particle case i.e. when $m_{Ab} = 0$ (so that $\mu_{\mathrm{in}} = 0$). In our case $\epsilon \simeq 0.036$. 


As $\bar{\omega}_{\mathrm{in}}$ changes, $e_{\mathrm{in}}$ and $i_{\mathrm{mut}}$ change so as to keep $H_{\mathrm{quad}}$ and $J_{\mathrm{tot}}$ constant. Armed with Eqs. \ref{eq:H_quad_covariant} and \ref{eq:J_tot_conservation}, we can then solve for all values of $e_{\mathrm{in}}$ (and $i_{\mathrm{mut}}$) that the inner binary will undergo. 
In practice, we compute a 2d grid of $e_{\mathrm{in}}$ and $\omega_{\mathrm{in}}$ ($i_{\mathrm{mut}}$ follows from $e_{\mathrm{in}}$ through Eq. \ref{eq:J_tot_conservation}) and plot the contour line that satisfies $H_{\mathrm{quad}} = H_{\mathrm{quad},0}$, where  $H_{\mathrm{quad},0}$ is computed using the measured orbital parameters. 

The effects of additional precession mechanisms in the inner binary can be easily included by adding their corresponding orbit-averaged Hamiltonian terms to Eq. \ref{eq:H_quad_covariant}. The extra Hamiltonian terms can be found for e.g. in \cite{Fabrycky07}. GR contributes an extra term 

\begin{align}
H_{\mathrm{GR}} = -\frac{3 G^2 m_{Aa} m_{Ab} (m_{Aa}+m_{Ab})}{a_{\mathrm{in}}^2 c^2} \frac{1}{(1-e_{\mathrm{in}}^2)^{1/2}}
\end{align}

Tidal precession contributes a term 

\begin{align}
H_{\mathrm{tide}} = -\frac{G}{8 a_{\mathrm{in}}^6} (m_{Ab}^2 k_{Aa} R_{Aa}^5 + m_{Aa}^2 k_{Ab} R_{Ab}^5) \frac{8 + 24 e_{\mathrm{in}}^2 + 3e_{\mathrm{in}}^4}{(1-e_{\mathrm{in}}^2)^{9/2}}
\end{align}

Precession due to the rotational bulge of the stars is more complex since it depends on both the star's angular rotational velocity $\Omega_j$ as well as on the inclination between the orbital and the spin axes $i_j$: 

\begin{align}
\begin{split}
&H_{\mathrm{rotate}} = - \frac{m_{Aa} m_{Ab}}{6 a_{\mathrm{in}}^3 (1-e_{\mathrm{in}}^2)^{3/2}} \\ &\left (\frac{k_{Aa} R_{Aa}^5}{m_{Aa}} \Omega_{Aa}^2 \times (3 \cos^2 i_{Aa} -1) + \frac{k_{Ab} R_{Ab}^5}{m_{Ab}} \Omega_{Ab}^2 \times (3 \cos^2 i_{Ab} -1) \right )  
\end{split} 
\end{align}

\noindent In general, $i_j$ changes with time since the orbital axis of the inner binary changes due to the ZKL oscillations as well as due to ``non-ZKL'' secular precession (about the total angular momentum axis) induced by the outer star B with a timescale of a few $\tau_{\mathrm{ZKL}}$. In general, for cases in which rotational bulge precession is important this would require to trace the $i_j$ in time and would break the validity of our analytical method. However, there is a further effect that can (and in the case of $\lambda$ Oph does) keep the $i_j$ effectively constant in time and ensures that our analytical method continues to be an excellent approximation. This is discussed next. 

\subsection{Slaved precession of the stellar spin axes}
\label{subsec:slaved_precession}

Companion Ab also applies a torque on Aa when $i_{Aa} \neq 0$, causing it to precess around the total angular momentum vector of the inner binary (and similarly for Aa on Ab). In the context of close binaries, this is often referred to as ``slaved'' precession \citep[perhaps the most famous example being the disk precession in SS433; e.g.][]{Waisberg19} and it is retrograde relative to the binary. If two conditions are met, namely (i) this precession timescale $\tau_{\mathrm{slaved}} \ll \tau_{\mathrm{ZKL}}$, and (ii) the total angular momentum of the inner binary is dominated by the orbital angular momentum, it follows that the stellar spin axes should effectively trace the inner orbital axis since they are constantly precessing around it, resulting in constant $i_j$ to a very good approximation. 

Regarding condition (ii), in the case of $\lambda$ Oph 

\begin{align}
J_{\mathrm{in}} = 7.25 \times 10^{45} \text{ J s} 
\end{align}

and the spin angular momenta 

\begin{align}
J_j = I_j \omega_j = \beta_j^2 m_j R_j v_j
\end{align}

\noindent where $I_j = \beta_j^2 m_j R_j^2$ is the moment of inertia, $\beta$ is the gyration radius and $v_j$ is the rotational velocity. For $\beta_{Aa} \simeq 0.188$, $\beta_{Ab} \simeq 0.204$ \citep[obtained from ][as were the apsidal constants]{Claret23} we find 

\begin{align}
J_{Aa} \simeq 4.1 \times 10^{43} \left ( \frac{v_{Aa}}{140 \text{ km}\text{ s}^{-1}} \right ) \text{ J s} \\
J_{Ab} \simeq 1.9 \times 10^{43} \left ( \frac{v_{Ab}}{140 \text{ km}\text{ s}^{-1}} \right ) \text{ J s}
\end{align}

\noindent so that if $v_{Aa} \sim v_{Ab} \sim v_A \sin i \simeq 140 \text{ km}\text{ s}^{-1}$ (see Section \ref{subsec:rv} for evidence that both Aa and Ab have high rotational velocities) we have $\frac{J_{Aa}+J_{Ab}}{J_{\mathrm{in}}} \simeq 0.008$. In fact, even if the stars were rotating at their critical speeds $v_{c,j} = \left ( \frac{G m_j}{R_j} \right )^{1/2}$, namely $v_{c,Aa} \simeq 428 \text{ km}\text{ s}^{-1}$ and $v_{c,Ab} \simeq 451 \text{ km}\text{ s}^{-1}$, we would have $\frac{J_{Aa}+J_{Ab}}{J_{\mathrm{in}}} \simeq 0.025$. Therefore condition (ii) is definitely satisfied. 

Regarding condition (i), we recall (see Appendix \ref{app:tau_slaved_derivation} for a derivation) that the slaved precession rate $\Omega_{\mathrm{slaved}}$ for Aa due to Ab is given by 

\begin{align} \label{eq:precRot}
\begin{split}
&\Omega_{\mathrm{slaved,Aa}} = \frac{2 \pi}{\tau_{\mathrm{slaved,Aa}}} = \\ &\frac{3}{2} \frac{Q_{Aa}}{I_{Aa}} \cos(i_{Aa}) \frac{\left( \frac{2 \pi}{P_{\mathrm{in}}} \right )^2}{\Omega_{Aa}} \frac{m_{Ab}}{m_{Aa} + m_{Ab}} \frac{1}{(1-e_{\mathrm{in}}^2)^{3/2}}
\end{split}
\end{align}

\noindent where $Q_{Aa}$ is the quadrupolar moment of Aa. In the case that $Q$ is dominated by the rotational bulge 

\begin{align} \label{eq:precRotQ}
Q_{Aa} = \frac{2}{3} k_{Aa} \Omega_{Aa}^2 R_{Aa}^5 \frac{1}{G}
\end{align}

\noindent so that 

\begin{align} \label{eq:tau_slaved}
\tau_{\mathrm{slaved,Aa}} = \frac{\beta_{Aa}^2}{k_{Aa}} \frac{G (m_{Aa} + m_{Ab}) P_{\mathrm{in}}^2 }{\Omega_{Aa} 2\pi R_{Aa}^3 \cos i_{Aa}} \frac{m_{Aa}}{m_{Ab}} (1-e_{\mathrm{in}}^2)^{3/2}
\end{align}

\noindent and similarly for Ab. Substituing the appropriate values for $\lambda$ Oph we find 

\begin{align}
\begin{split}
&\tau_{\mathrm{slaved,Aa}} \approx \tau_{\mathrm{slaved,Ab}} \simeq 1.0 \text{ kyr} \left ( \frac{v_j}{140 \text{ km}\text{ s}^{-1}} \right )^{-1} \\ &\times \left ( \frac{\cos i_j}{\cos \pi/3} \right )^{-1} \frac{(1-e_{\mathrm{in}}^2)^{3/2}}{(1-0.71^2)^{3/2}} 
\end{split}
\end{align}

\noindent where $v_j$ is the rotational velocity. Therefore, condition (i) is also satisfied except for $i_j$ that are extremely close to $90^{\degr}$ (for $v_j = 140 \text{ km}\text{ s}^{-1}$ and $e_{\mathrm{in}}=0.71$ we have $\tau_{\mathrm{slaved,Aa}} \approx \tau_{\mathrm{slaved,Ab}} = \tau_{\mathrm{ZKL}} \simeq 130 \text{ kyr}$ for $i_j \approx 89.8^{\degr} $). 

In Appendix \ref{app:numerical_solution} we compare our analytical solution assuming constant stellar spin axes inclination angles $i_j$ with a numerical solution obtained by integrating the equations of motion for different initial values of $i_j$. As expected, the comparison confirms our conclusion that in the case of $\lambda$ Oph the slaved precession is fast enough as to render the analytical solution an excellent approximation, unless the initial inclination angle is very close to $90^{\degr}$ (namely $i_j \gtrsim 89^{\degr}$). 

\subsection{Dynamics in Lambda Ophiuchi}
\label{subsec:dynamics}

\begin{figure*}[]
\centering
\includegraphics[width=0.8\textwidth]{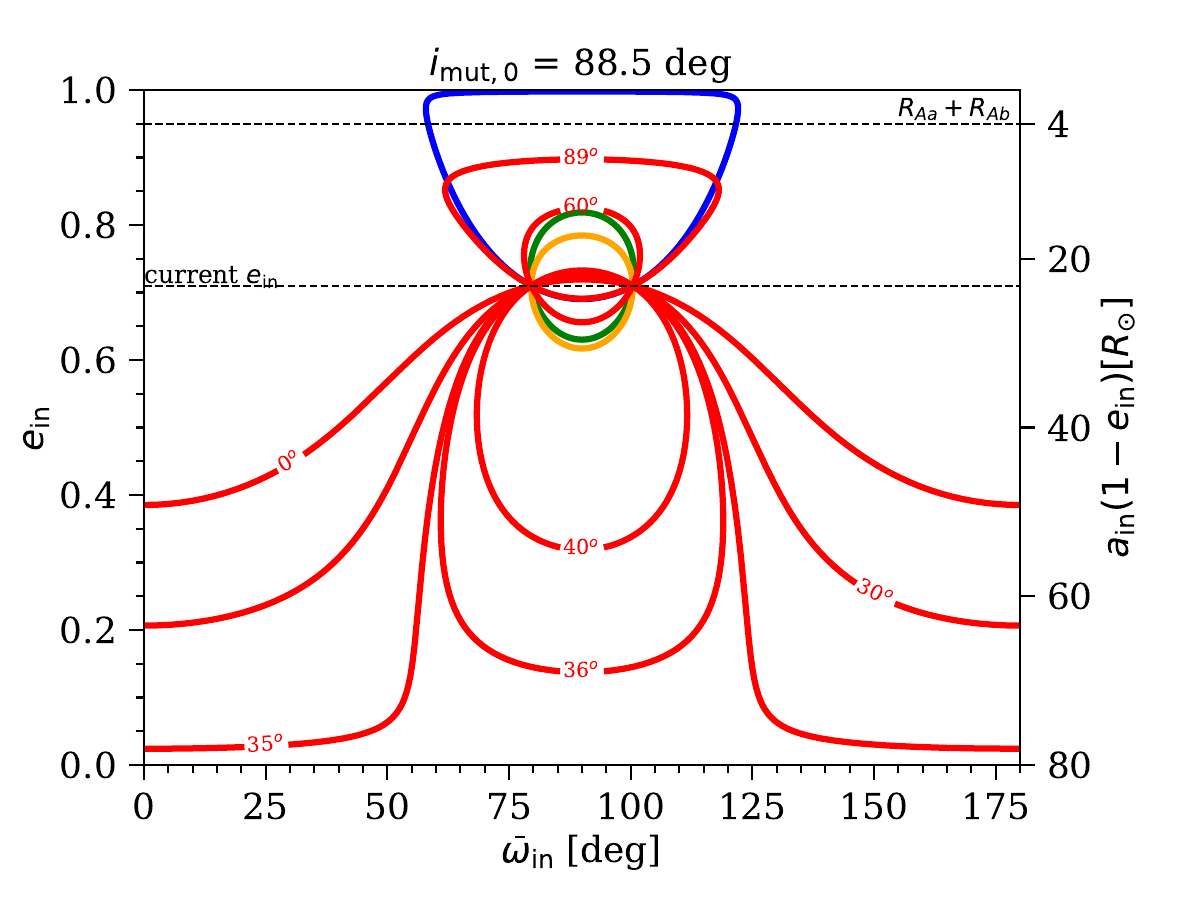} \\
\includegraphics[width=0.8\textwidth]{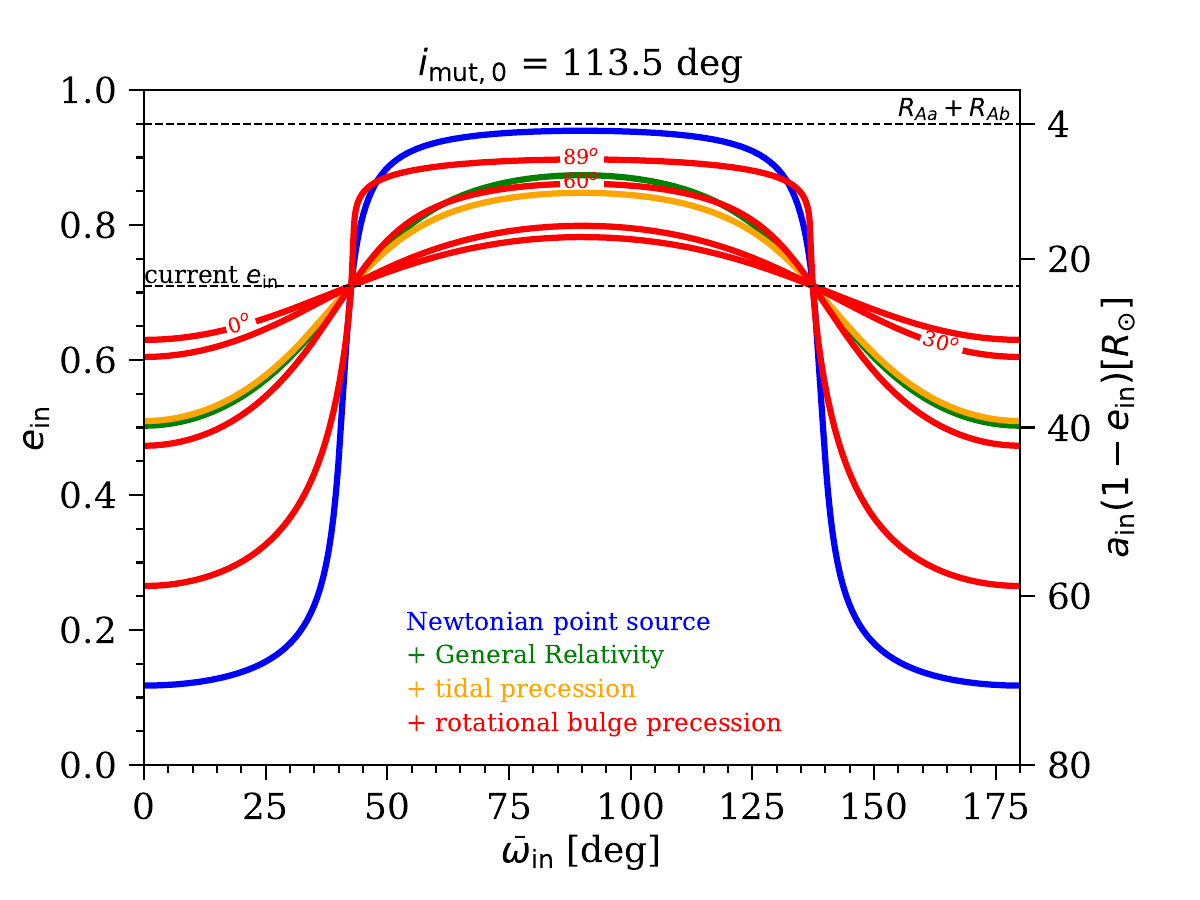}
\caption{\label{fig:e_vs_omega} von Zeipel-Kozai-Lidov oscillations for the inner binary in Lambda Ophiuchi for the case in which the current mutual inclination is $88.5^{\degr}$ (top) and $113.5^{\degr}$ (bottom). The labels for the red lines refer to the (currently unknown) inclination between the stellar spin axes and the inner orbital axis.}
\end{figure*}

Figure \ref{fig:e_vs_omega} shows the resulting dynamics in the inner binary of Lamdba Ophiuchi for the two possible current mutual inclinations\footnote{Note that due to the $\sin^2 \bar{\omega}_{\mathrm{in}}$ dependence in the quadrupolar Hamiltonian, the dynamical lines are symmetric around $\bar{\omega}_{\mathrm{in}}=\frac{\pi}{2}$ and repeat for $\pi \leq \bar{\omega}_{\mathrm{in}} \leq 2\pi$.}. We added each contribution to the Hamiltonian progressively so that their corresponding effect on the dynamics can be clearly gauged. For the rotational bulge contribution, we plot the resulting curves for stellar spin axes inclinations relative to the inner orbital axis of $i_{Aa} = i_{Ab} = i_A = 0^{\degr}$, $30^{\degr}$, $60^{\degr}$ and $89^{\degr}$ (assumed to the same for both stars)\footnote{The primary Aa dominates the rotational bulge precession due to its larger radius, so that the dynamics depends only weakly on $i_{Ab}$ (see Figure \ref{fig:delta_e_2d}), unless the secondary Ab happens to have a much higher rotational velocity.}. The rotational velocity of Aa was fixed to $v_{Aa} = 140 \text{ km}\text{ s}^{-1}$ based on the measured $v \sin i$ and that of the secondary to $v_{Ab} = v_{Aa} \frac{R_{Ab}}{R_{Aa}} = 84 \text{ km}\text{ s}^{-1}$ so that they have the same angular velocity\footnote{More extreme curves are possible for higher rotational velocities but the real rotational velocities are unlikely to be much higher than what we assumed since there is no evidence for a decretion disk in the system such as shell absorption lines, emission lines and near-infrared excess.}. 

It is interesting to note that the pure quadrupolar potential (blue curve) would result in the two stars colliding in the case $i_{\mathrm{mut,0}}=88.5^{\degr}$ and almost graze in the case $i_{\mathrm{mut,0}}=113.5^{\degr}$ (the right y-axis shows the pericenter distance with $R_{Aa}+R_{Ab} \simeq 4 R_{\odot}$ marked). As expected, the precession terms due to General Relativity (green curve) and to a lesser extent due to the tidal bulges (orange curve) play a crucial role in modulating the ZKL oscillations and reducing the maximum eccentricity $e_{\mathrm{max}}$ reached. On the other hand, the rotational bulge precession (red curves) has a dominant effect and different values of $i_A$ can lead to very diverse dynamical behavior. For $i_{\mathrm{mut}}=88.5^{\degr}$, $i_A \leq 35^{\degr}$ results in circulation whereas $i_A > 35^{\degr}$ results in libration; for $i_{\mathrm{mut}}=113.5^{\degr}$, the ZKL oscillations take the form of circulation for any $i_A$. 

If the spin axes are currently aligned with the inner orbit ($i_{A}=0^{\degr}$), then for $i_{\mathrm{mut,0}}=88.5^{\degr}$ the eccentricity oscillates between $e_{\mathrm{min}}=0.38$ and $e_{\mathrm{max}}=0.72$ (oscillation amplitude $\Delta e = 0.33$); in this case, the inner binary is currently very close to its maximum eccentricity. For $i_{\mathrm{mut,0}}=113.5^{\degr}$ and $i_{A}=0^{\degr}$, the corresponding values are $e_{\mathrm{min}}=0.63$ and $e_{\mathrm{max}}=0.78$ ($\Delta e = 0.15$) and the inner binary is currently near the midpoint of the oscillation. However, given that the stellar spins are currently not synchronized (i.e. $\frac{R}{v \sin i} \ll P_{\mathrm{in}}$), there is no strong reason to expect that $i_{A}$ is small. 

\begin{figure}[]
\centering
\includegraphics[width=0.5\textwidth]{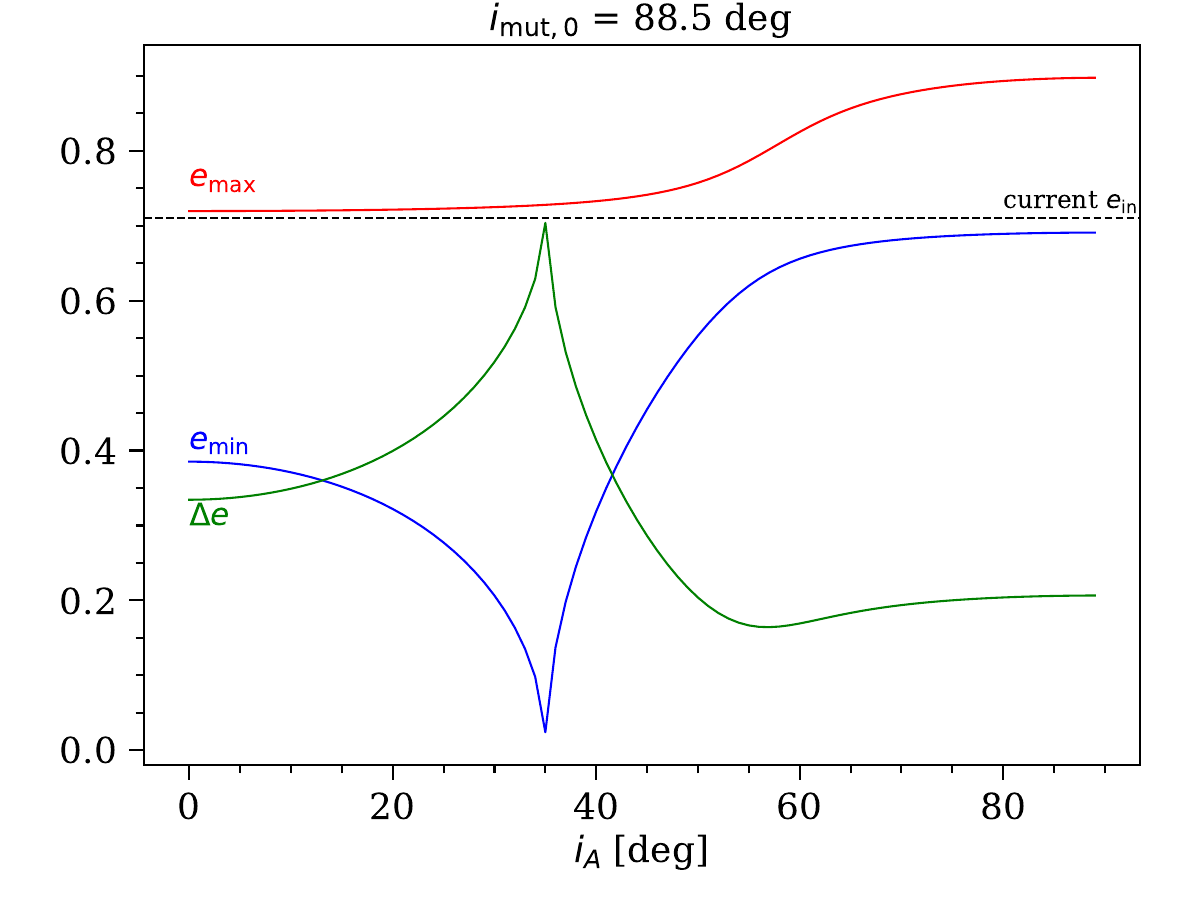} \\
\includegraphics[width=0.5\textwidth]{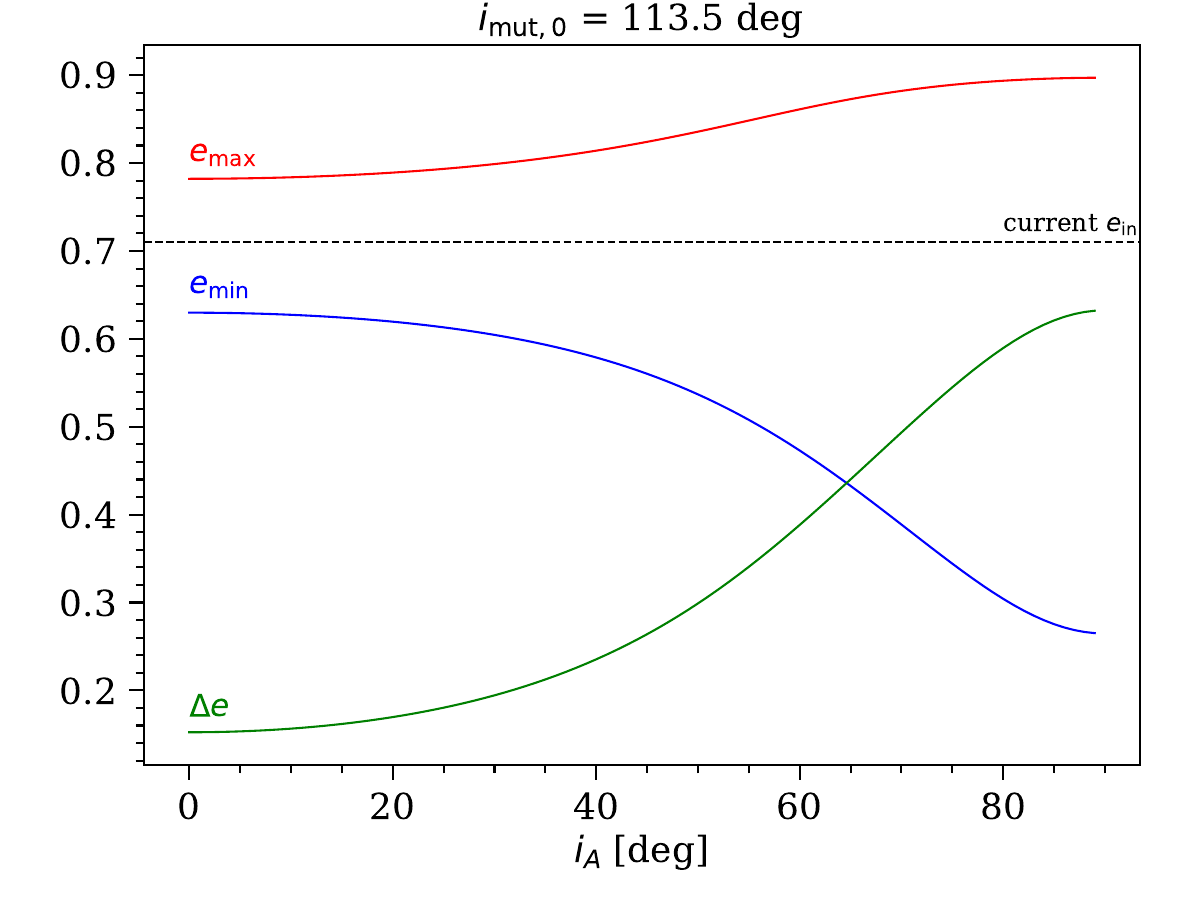}
\caption{\label{fig:emax_emin_delta_e} Maximum eccentricity ($e_{\mathrm{max}}$), minimum eccentricity ($e_{\mathrm{min}}$) and amplitude ($\Delta e = e_{\mathrm{max}} - e_{\mathrm{min}}$) in the von Zeipel-Kozai-Lidov oscillations of Lambda Ophiuchi Aa+Ab as a function of the spin axes misalignment relative to the inner orbital axis ($i_A$). The two plots correspond to the two possible current mutual inclinations of the inner and outer orbits.}
\end{figure}

In Figure \ref{fig:emax_emin_delta_e}, we plot $e_{\mathrm{max}}$, $e_{\mathrm{min}}$ and $\Delta e$ as a function of $i_A$ for the two possible mutual inclinations. Corresponding two-dimensional plots as a function of both $i_{Aa}$ and $i_{Ab}$ (i.e. without assuming $i_{Aa} = i_{Ab}$) can be found in Appendix \ref{app:delta_e_2d}, but since the precession is dominated by the Aa term they do not differ significantly from Figure \ref{fig:emax_emin_delta_e}. For $i_{\mathrm{mut,0}}=113.5^{\degr}$, the minimum $e_{\mathrm{max}} =0.78$ and $\Delta e = 0.15$ happen for $i_A = 0^{\degr}$ and the maximum $e_{\mathrm{max}} =0.90$ and $\Delta e = 0.63$ happen for $i_{A}=90^{\degr}$. For $i_{\mathrm{mut,0}}=88.5^{\degr}$, $e_{\mathrm{min}}$ has a more complex behavior so that the maximum $\Delta e = 0.70$ is reached for $i_{A} = 35^{\degr}$ and the minimum $\Delta e = 0.16$ for $i_{A} = 57^{\degr}$; the maximum $e_{\mathrm{max}} =0.90$ is reached for $i_{A} = 90^{\degr}$ and the minimum $e_{\mathrm{max}} =0.72$ for $i_{A} = 0^{\degr}$. We refrain from quoting probabilities for $\Delta e$ because the probability distribution for $i_A$ is unknown; for instance, there is no reason why it should be isotropic ($p(i_A) d i_a  = \sin i_A d i_A$).

\section{Discussion}
\label{sec:discussion}

\subsection{Formation and future evolution}
\label{sec:tidal_dissipation}

It is interesting to consider what could have caused a relatively compact triple system ($a_{\mathrm{out}}=46 \text{ au}$) to have an inner orbit that is misaligned by close to $90^{\degr}$ relative to the outer orbit. It is possible that the triple system was formed dynamically through a close encounter in its parent cluster rather than from the collapse of a single molecular core. Alternatively, the misalignment may have cascaded from the very wide companion C (see \ref{subsec:wide_companion}). 

One may wonder whether the ZKL oscillations have played a role in the formation of the close inner binary by inducing tidal dissipation and whether such migration is currently happening. We argue that this is not likely to be the case because the large projected rotational velocity $v \sin i \simeq 140 \text{ km}\text{ s}^{-1}$ (see \ref{subsec:rv} for evidence that the secondary also has a high $v \sin i$) implies a stellar rotation period $P_{\mathrm{rot,Aa}} \lesssim 22 \text{ hr}$ that is much shorter than the pseudo-synchronization period, which for the current eccentricity $e=0.71$ is $P_{\mathrm{pseudo-sync}} \simeq 6 \text{ day}$ \citep{Hut81}\footnote{For a system undergoing high-amplitude ZKL oscillations an ``effective eccentricity'' should be used for estimating the pseudo-synchronization period but as can be seen in Figure \ref{fig:emax_emin_delta_e} this should not be higher than the current $e$ in this case.}. 

A basic feature of tidal dissipation theories is that synchronization should happen much faster than migration \citep[e.g.][]{Zahn77,Hut81}. For example, in the case of dynamical tides (appropriate for this case since the stars have a radiative envelope):

\begin{align}
\left | \frac{a}{\frac{da}{dt}} \right | \sim \left ( \frac{a}{R} \right )^{21/2} \\
\left | \frac{\Omega}{\frac{d\Omega}{dt}} \right | \sim \left ( \frac{a}{R} \right )^{17/2}
\end{align}

\noindent so that synchronization should happen very roughly $\left ( \frac{a_{\mathrm{in}}}{R_{Aa}} \right )^2 = 10^3$ times faster than migration. On the other hand, the maximum eccentricity that the binary could possibly reach is $e_{\mathrm{max}} \simeq 0.90$ (see \ref{subsec:dynamics}), corresponding to a pericenter distance $a_p = a_{\mathrm{in}}(1-e_{\mathrm{max}}) \simeq 0.037 \text{ au}$ so that $\frac{a_p}{R_{Aa}} \approx 3$ could be small enough to cause tidal synchronization and in this case such high $e_{\mathrm{max}}$ (and their corresponding $i_A$) could potentially be excluded. However, this would require a detailed calculation coupling tidal dissipation (whose strength is still rather uncertain in the case of dynamical tides) with the ZKL oscillations and is beyond the scope of this paper. 

As the primary evolves and expands, tidal dissipation and circularization will be inevitable. At this point, the ZKL eccentricity oscillations will play an important role since the effective ``average'' eccentricity $e_{\mathrm{in,avg}}$ will determine the semi-major axis after circularization. By conservation of angular momentum 

\begin{align}
a_{\mathrm{final}} = a_{\mathrm{in}} (1-e_{\mathrm{in,avg}}^2) 
\end{align}

\noindent where we have neglected the spin angular momentum since it is very small comparable to the orbital angular momentum (\ref{subsec:slaved_precession}). A small $e_{\mathrm{min}}$ during the ZKL oscillations will result in a larger $a_{\mathrm{final}}$, which would have important implications for the future evolution of the system once it inevitably undergoes common envelope evolution.

\subsection{The possible dynamical effect of the wide companion C}
\label{subsec:wide_companion}

The very wide $0.6 M_{\odot}$ companion C at a projected separation of about 6,400 au could also exert dynamical effects on the system depending on the mutual inclination between (A+B)+C and A+B and its orbital eccentricity. The corresponding Kozai-Lidov timescale for eccentricity and inclination oscillations in A+B is 

\begin{align}
\tau_{\mathrm{KL}} \sim 5.4 \text{ Gyr} \left ( \frac{a_{(A+B)+C}}{6400 \text{ au}} \right ) (1-e_{(A+B)+C}^2)^{3/2} 
\end{align}

Therefore, the dynamical influence of component C is only relevant if $e_{(A+B)+C}$ is very high (e.g. $\tau_{\mathrm{KL}} \sim 15 \text{ Myr}$ for $e_{(A+B)+C} = 0.99$) . In that case, one could imagine a scenario in which Aa+Ab and A+B were initially aligned and it was a misaligned C that caused the orbital axis of A+B to oscillate, cascading into the current oscillations in Aa+Ab \footnote{We note that precession terms that could potentially damp the KL oscillations are irrelevant for the (A+B)+C system.}. In any case, our predicted dynamical behavior for Aa+Ab should remain valid for very long timescales of the order of millions of years even if component C turns out to be dynamically important in the very long run.

\subsection{Prospects for breaking the $\Omega \leftrightarrow \omega$ degeneracies}
\label{subsec:rv}

In order to determine which of the two possible $i_{\mathrm{mut}}$ is the true one, it is necessary to break the $(\Omega, \omega) \leftrightarrow (\Omega + 180^{\degr}, \omega + 180^{\degr})$ degeneracy for both the inner and the outer orbits. For this it suffices to measure the sign of the change in radial velocity (RV) of one of the binary components between two epochs; a full radial velocity curve is not actually needed. However, the closeness of components A and B as well as the large $v \sin i$ can make RV measurements challenging. 

\begin{figure}[]
\centering
\includegraphics[width=0.5\textwidth]{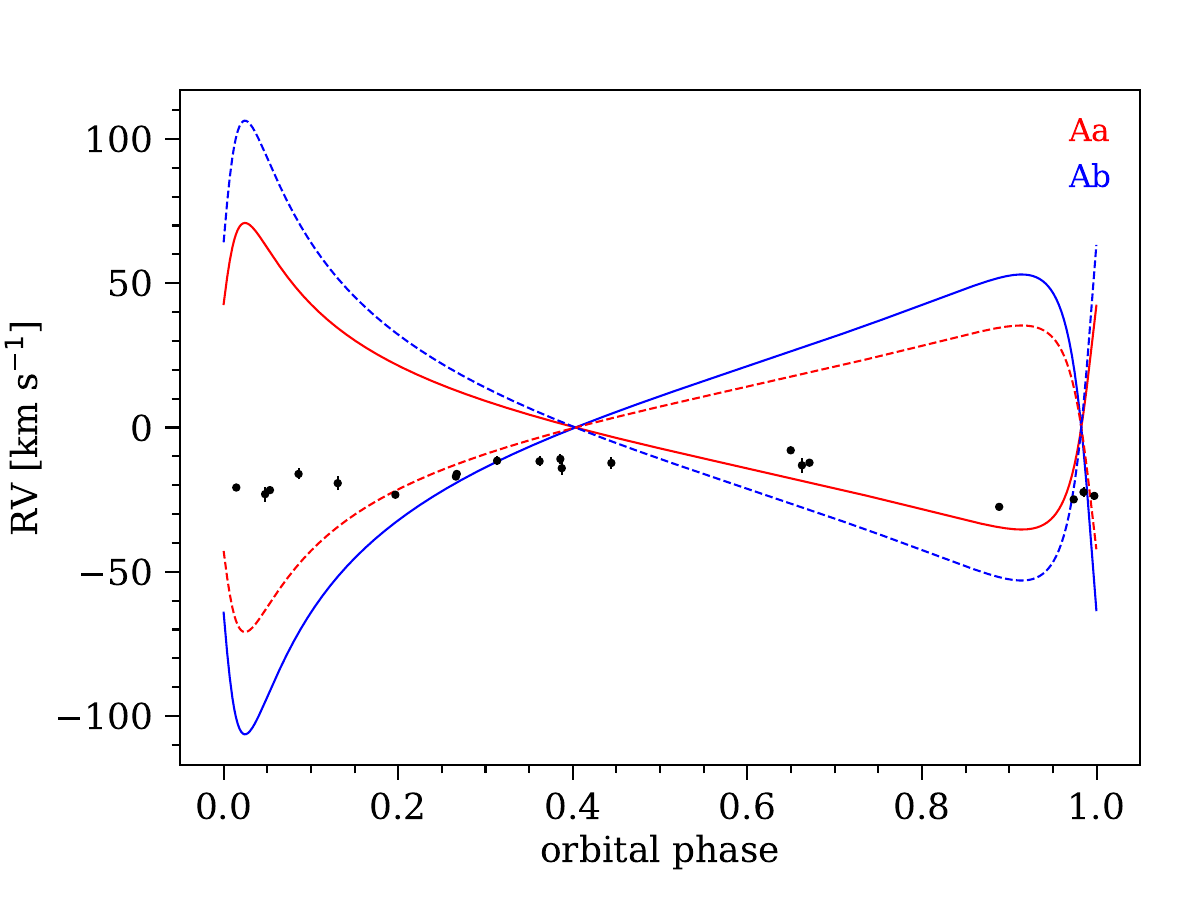}
\caption{\label{fig:rv_curve} Predicted radial velocity curves for components Aa and Ab (two possible solutions for $\omega$ in blue and $\omega+180^{\degr}$ in red) with a systemic velocity of zero. The reported measurements in \cite{Abt80} for the net spectrum are shown in black.}
\end{figure}

Figure \ref{fig:rv_curve} shows the predicted RV curves for the primary Aa and secondary Ab based on the orbital parameters we measured (two possibilities for $\omega$ in blue and $\omega+180^{\degr}$ in red are shown) and for a systemic velocity of zero. We also plot in black the reported RV measurements in \cite{Abt80} from the net spectrum of component A. Clearly the RV variations of the net spectrum are much smaller than that for each individual component, which can be explained if both components have an important contribution to the spectral lines so that the net RV is close to constant. In particular, this implies that both components should have a large $v \sin i$ as measured from the net spectrum. 

Components A and B are still well separated on sky (about 1.3" as of 2024) so that it should be possible to obtain separate spectra under good seeing conditions even without Adaptive Optics. Measuring the sign of the RV change of component B (a single star) over a few years -- the RV semi-amplitude of component B is about $3.5 \text{ km}\text{ s}^{-1}$ -- should break the degeneracy for the outer orbit. Meanwhile, a series of spectra of component A at high spectral resolution should allow for spectral disentangling of components Aa and Ab given their large RV amplitudes comparable to $v \sin i$, so that it should also be possible to break the degeneracy for the inner orbit. We conclude that determining the unique value of $i_{\mathrm{mut}}$ should be possible with straightforward albeit dedicated spectroscopic observations.

\subsection{Prospects for measuring the stellar spin axes orientation}

In order to measure the inclination of the spin axes of Aa and Ab relative to the inner binary it is necessary to measure both their inclination relative to the line of sight $i_{\mathrm{spin}}$ as well as their longitude of the ascending node $\Omega_{\mathrm{spin}}$. 

If there is rotational modulation due to surface inhomogeneities in either or both of the stars, it might be possible to measure $i_{\mathrm{spin}}$ from the rotation period combined with the radii and projected rotational velocities. However, the large brightness of $\lambda$ Oph requires instrumentation that can deal well with saturated targets such as the Transiting Exoplanet Survey Satellite \citep[TESS; ][]{Ricker15}. Unfortunately, there are currently no TESS observations of $\lambda$ Oph. 

Both $i_{\mathrm{spin}}$ and $\Omega_{\mathrm{spin}}$ could also be potentially measured by resolving the stellar disk with very high resolution interferometry since the fast rotation leads to stellar obliquity and gravity darkening. Such measurements have been done for a handful of nearby fast rotators such as $\alpha$ Aql = \textit{Altair} \citep[equatorial angular diameter $\theta_{\mathrm{eq}} = 3.7 \text{ mas}$; ][]{Monnier07}, $\beta$ Cas = \textit{Caph} \citep[$\theta_{\mathrm{eq}} = 2.1 \text{ mas}$; ][]{Che11} and $\alpha$ Leo = \textit{Regulus} \citep[$\theta_{\mathrm{eq}} = 1.6 \text{ mas}$; ][]{Che11}. The angular diameters of $\lambda$ Oph Aa ($\theta_{Aa} = 0.43 \text{ mas}$) and Ab ($\theta_{Ab} = 0.27 \text{ mas}$) are more challenging to resolve and would certainly require optical rather than near-infrared interferometry; even using the extended AT configuration our VLTI/GRAVITY observations are still limited to an angular resolution of about 2 mas. 

Finally, we note that there is a model-independent technique that can be used to constrain $\Omega_{\mathrm{spin}}$ using high-spectral resolution interferometry even when the formal resolution is a few times larger than the stellar disk. Namely, if $v \sin i$ is comparable to or larger than the spectral resolution, the rotation should create a typical S-shaped signature in the differential visibility phases across a spectral line, from which $\Omega_{\mathrm{spin}}$ can be easily extracted. We applied this to measure $\Omega_{\mathrm{spin}}$ from VLTI/GRAVITY observations of $\gamma$ Tra \citep[$v \sin i = 200 \text{ km}\text{ s}^{-1}$, $\theta = 1 \text{ mas}$][]{Waisberg24}. Our current VLTI/GRAVITY observations of $\lambda$ Oph do not show any statistically significant rotation signature in the differential phases, but a longer observation during a single night would achieve a better SNR and have great potential in measuring $\Omega_{\mathrm{spin}}$. 

\subsection{Alternative methods for constraining triple dynamics}

In addition to the combination of historical outer orbits and interferometric inner orbits in nearby multiple systems, there are other possible methods for solving the dynamics in multiple systems. For instance, space astrometry with Gaia may be able to provide for an astrometric orbit of the inner binary. Ideally, such orbits should still be validated through radial velocity measurements, which are also useful for converting the photocenter semi-major axis to the actual one. A combination of outer orbits from ORB6 and inner orbits from Gaia therefore seems promising for dynamical studies. However, a drawback of this approach is that very often Gaia has significant trouble with nearby visual binaries, which typically have orbits of size of about 0.1-1" comparable to Gaia's Point Spread Function. Furthermore, Gaia DR3 astrometric solutions are notoriously lacking for stars of about $2 M_{\odot}$ and above and so this approach currently is not very promising for stars with intermediate mass and above but this may change in future data releases. 

It is also possible to solve for the dynamics in multiple systems in case the inner binary happens to an eclipsing binary. The outer companion induces eclipse-depth variations, which can be used to derive relevant orbital parameters for both the inner and outer orbits including their mutual inclination. This has been done for example in \citep{Borkovits22}, who report on two systems (KIC 5731312 and KIC 8023317) in which the inner eclipsing binaries are inferred to be currently undergoing Kozai-Lidov oscillations of relatively large amplitude $\Delta e \sim 0.3$. Although the \textit{Kepler} and \textit{TESS} missions have allowed for the discovery of a continuously increasing number of eclipsing binaries in compact multiple systems, such configurations are still very rare and the vast majority of such systems are of low mass. Such studies are also restricted to (i) close (eclipsing) binaries for which apsidal precession in the inner binary is likely to significantly damp ZKL oscillations, and (ii) low hierarchy configurations for which the inner orbital plane precession (leading to the eclipse depth variations) can be measured on human timescales. Therefore, eclipsing binaries in multiple systems are unlikely to provide strong constraints on the orbital architectures and the relevance of ZKL oscillations for systems of intermediate-mass and above. 

\subsection{Conclusion}
\label{sec:conclusion}

In this paper we have solved for the orbital architecture of the hierarchical triple system within the nearby and bright intermediate-mass star Lambda Ophiuchi. The outer orbit A+B with a period of 129.5 yrs was solved based on astrometric observations spanning the period 1825-2024 collected in ORB6, while the inner orbit with a period of 42 days was solved based on our new VLTI/GRAVITY interferometric observations. We found that the orbits are significantly misaligned with a mutual inclination that is either $88.5\pm1.9^{\degr}$ or $113.5\pm1.9^{\degr}$. This results in von Zeipel-Kozai-Lidov (ZKL) oscillations being a crucial component to the dynamics of the system. While there have been a scarce number of triple system solutions for which ZKL oscillations have been estimated in the literature, we are not aware of \textit{any} previous case with such a high mutual inclination and the potential for high amplitude eccentricity oscillations. 

We calculated the dynamics of the system using the quadrupolar approximation and taking into account the relevant apsidal precession terms in the inner binary due to General Relativity, tidal bulges and rotational bulges. In particular, we showed that in the case of $\lambda$ Oph the inclination between the stellar spin axes and the inner orbital axis remains approximately constant due to ``slaved'' precession in the inner binary, which results in simplified dynamics that can be solved analytically based on conservation of energy and total angular momentum without the need to integrate the equations of motion. 

The resulting dynamics are rather interesting. The additional precession terms are crucial to modulate the ZKL oscillations (which would otherwise cause the inner binary to quickly merge). Furthermore, the rotational bulge precession has a dominant role so that the dynamics of the system strongly depends on the currently unknown inclination angle between the stellar spin axes and the orbital angular momentum vector. The amplitude of the ZKL oscillations is at least $\Delta e \simeq 0.15$ and can be as high as $\Delta e \simeq 0.70$. We argued that further spectroscopic observations can determine which of the two possible current mutual inclination angles is the correct one, while further interferometric observations have the potential to measure the orientation of the stellar spin axes. Therefore, there is a good chance that the dynamics of the system can be completely solved for.

$\lambda$ Oph shows that there is a great potential in combining historical astrometric observations of outer orbits with interferometric observations of inner orbits of nearby multiple systems in order to solve for their dynamics. Expanding this technique to further multiple systems is necessary in order to inform our knowledge of the orbital architectures of multiple systems with stars of intermediate-mass and above and assess the role that ZKL oscillations play in their evolution. 

\section*{Acknowledgments}

This research has made use of the CDS Astronomical Databases SIMBAD and VIZIER, NASA's Astrophysics Data System Bibliographic Services, NumPy \citep{van2011numpy} and matplotlib, a Python library for publication quality graphics \citep{Hunter2007}.

\section*{Data availability}
The VLTI/GRAVITY data underlying this article is publicly available from the ESO archive. Reduced data can be provided by the author on request.

\bibliographystyle{aasjournal}
\bibliography{main}{}

\appendix

\section{A. Parameter distributions for orbital fits}
\label{app:orbital_params_distributions}

\begin{figure*}[]
\centering
\includegraphics[width=\textwidth]{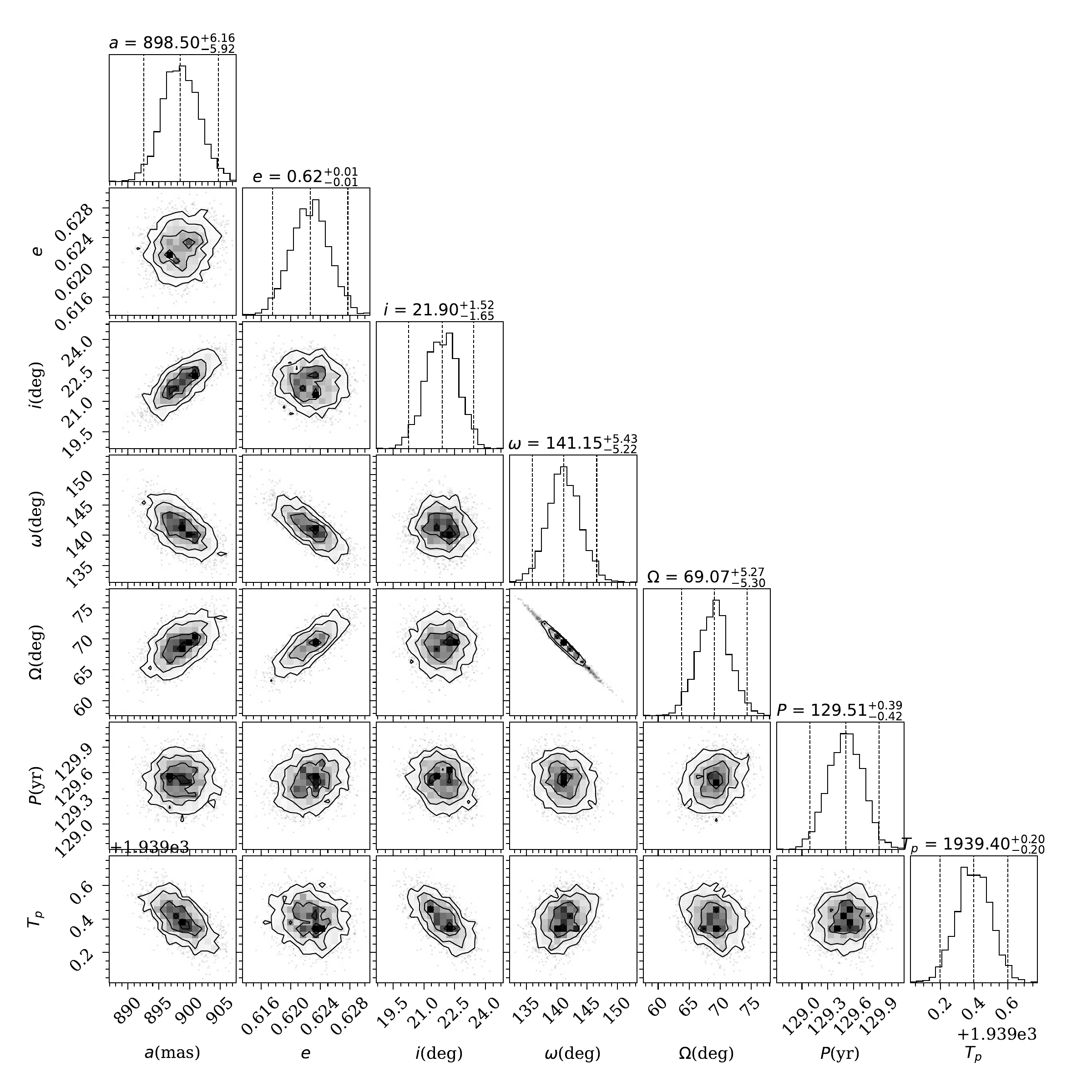}
\caption{\label{fig:corner_A+B} Orbital parameters distributions for the A+B outer orbit.}
\end{figure*}

\begin{figure*}[]
\centering
\includegraphics[width=\textwidth]{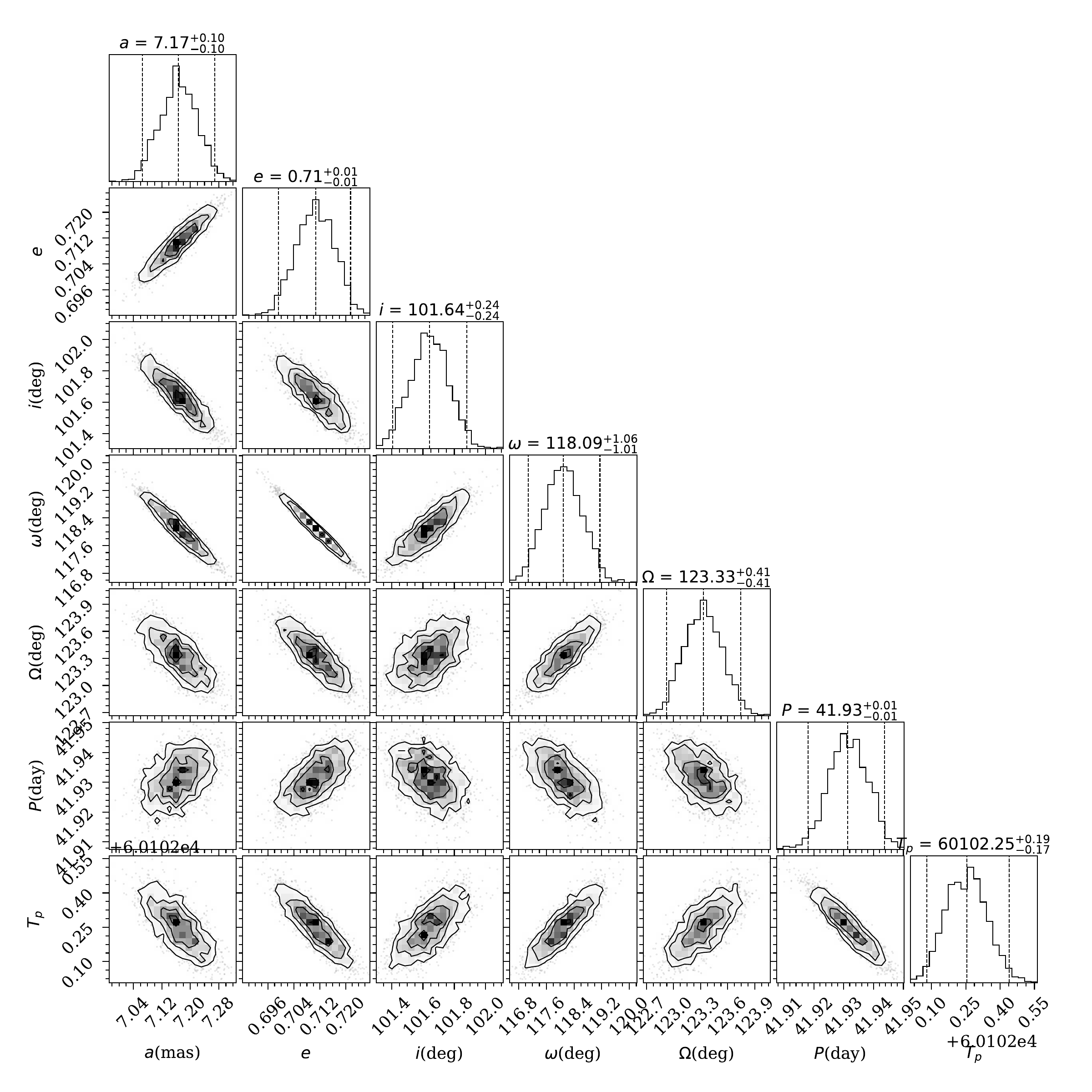}
\caption{\label{fig:corner_Aa+Ab} Orbital parameters fit distributions for the Aa+Ab inner orbit.}
\end{figure*}

The full distributions of best fit orbital parameters for the A+B and Aa+Ab orbits of Lambda Ophiuchi are shown in Figures \ref{fig:corner_A+B} and \ref{fig:corner_Aa+Ab}.

\section{B. VLTI/GRAVITY observations and best-fit parameters}
\label{app:gravity_observations}

Table \ref{table:observations_gravity} shows details of the VLTI/GRAVITY observations of $\lambda$ Oph A as well as the best-fit binary model parameters for Aa+Ab for each epoch. 

\begin{table*}
\centering
\caption{\label{table:observations_gravity} VLTI/GRAVITY observations of Lambda Ophiuchi A and best-fit binary model parameters for A$a$+A$b$.}
\begin{tabular}{ccccccc}
\hline \hline
\shortstack{date\\MJD} & \shortstack{seeing\\@ 500 nm (")} & \shortstack{AT configuration} & \shortstack{$B_{\mathrm{proj,max}}$ (m) \\$\theta_{\mathrm{max}}$ (mas)} & \shortstack{$\frac{f_{\mathrm{A_b}}}{f_{\mathrm{A_a}}}$ (\%)\\K band} & \shortstack{$\Delta \alpha_*$\\(mas)} & \shortstack{$\Delta \delta$\\(mas)} \\ [0.3cm]

\shortstack{2023-05-17\\60081.210} & 1.0 & K0-G2-D0-J3 & \shortstack{94.4\\4.8} & 32.9 & 3.598 & -4.975 \\ [0.3cm]

\shortstack{2024-05-19\\60449.223} & 1.1-1.3 & A0-B2-D0-J3 & \shortstack{128.3\\3.5} & 32.9 & -0.172 & -2.404 \\ [0.3cm]

\shortstack{2024-06-17\\60478.168} & 1.0 & K0-B5-D0-J6 & \shortstack{186.1\\2.4} & 32.6 & 2.226 & -0.991 \\ [0.3cm]

\shortstack{2024-08-11\\60533.031} & 1.0 & K0-G1-D0-J3 & \shortstack{120.9\\3.8} & 33.6 & -0.224 & -2.358 \\ [0.3cm]

\shortstack{2024-08-14\\60536.094} & 0.6 & A0-B5-J2-J6 & \shortstack{198.7\\2.3} & 32.3 & 1.127 & -3.413 \\ [0.3cm]

\shortstack{2024-09-03\\60556.002} & 0.5 & A0-G1-J2-K0 & \shortstack{122.5\\3.7} & 33.6 & 5.875 & -4.763 \\ [0.3cm]

\hline
\end{tabular}
\tablenotetext{0}{Notes:}
\tablenotetext{0}{The estimated astrometric error for each observation is 0.016 mas.}
\end{table*}

\begin{figure*}[]
\centering
\includegraphics[width=\textwidth]{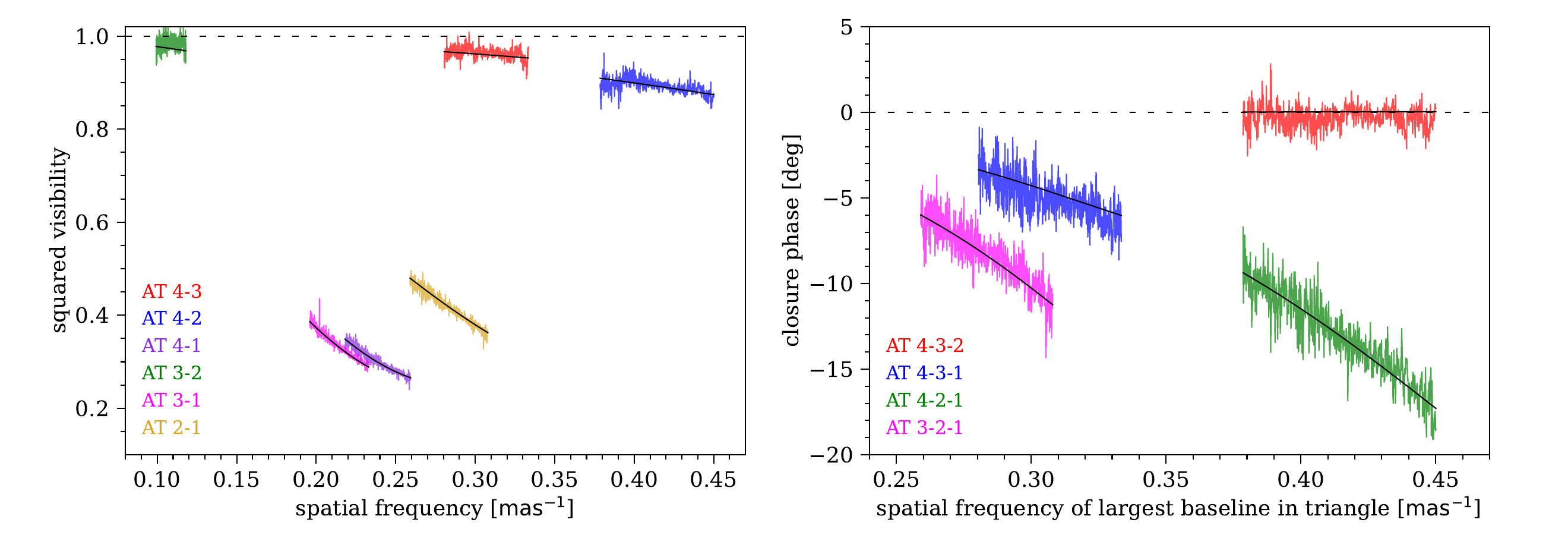}\\
\includegraphics[width=\textwidth]{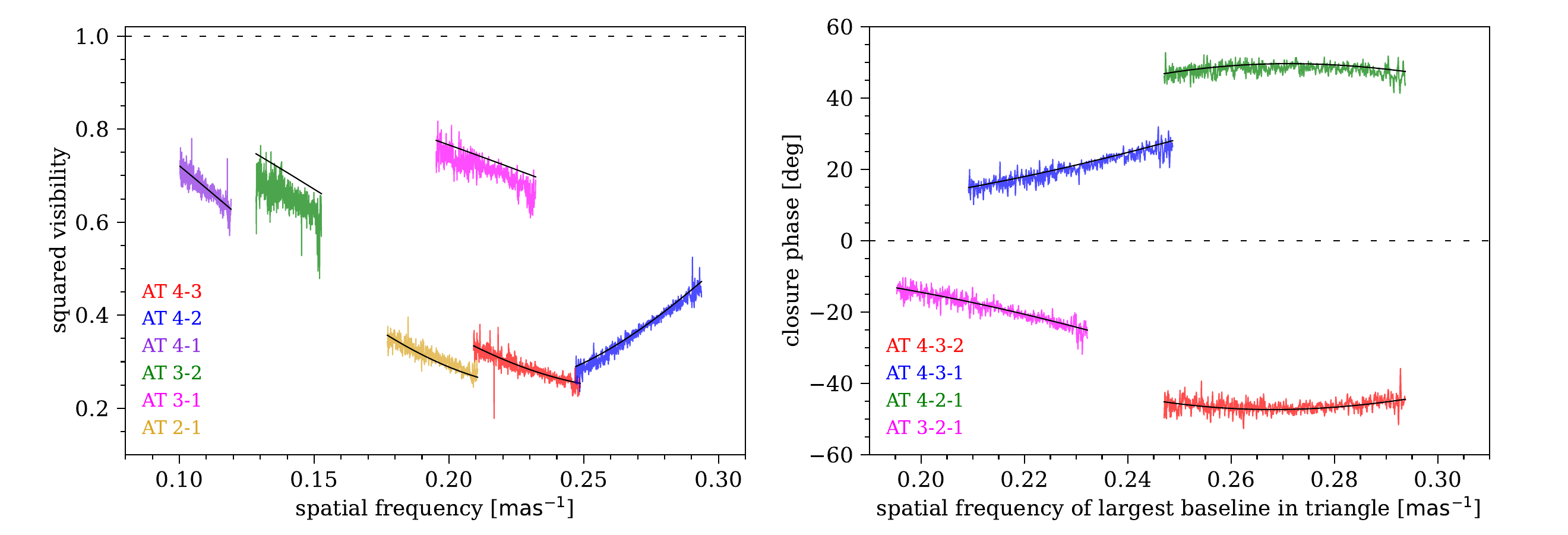}\\
\includegraphics[width=\textwidth]{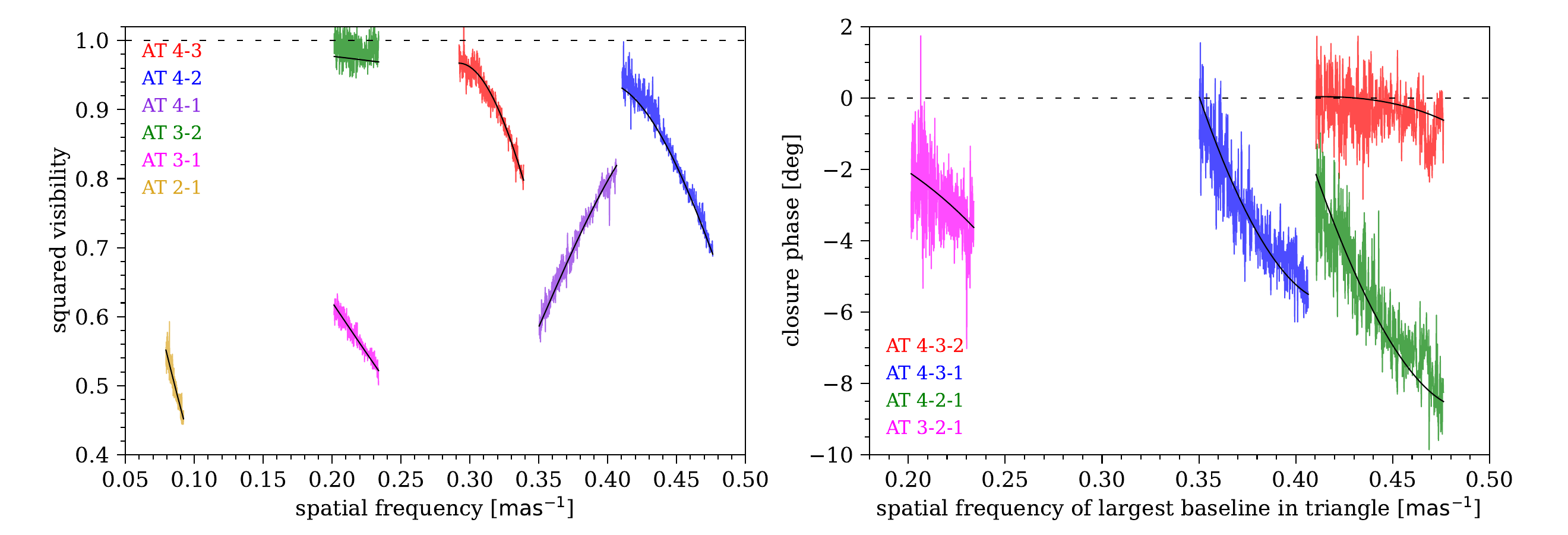} \\
\includegraphics[width=\textwidth]{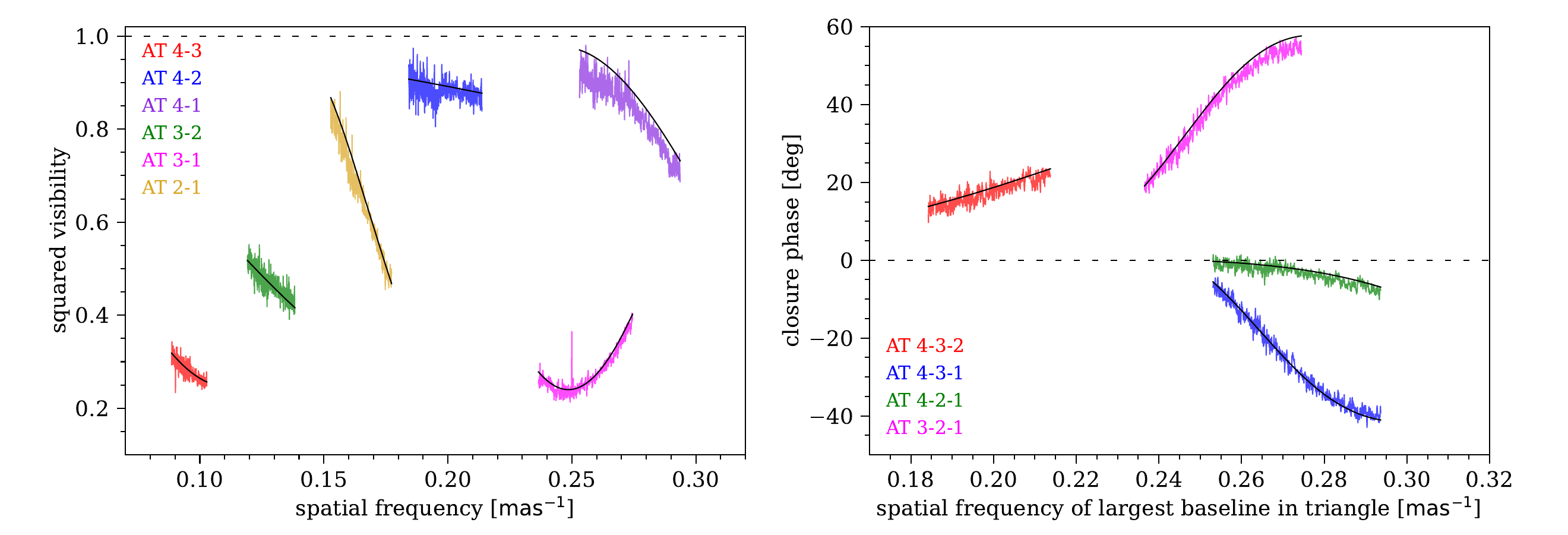} 
\caption{\label{fig:gravity_fits_additional} VLTI/GRAVITY data (colored) for $\lambda$ Oph Aa+Ab and best fit binary model (solid black) for Epochs 2024-06-17, 2024-08-11, 2024-08-14 and 2024-09-03. The dashed lines show the expected values for a single unresolved star.}
\end{figure*}

Figure \ref{fig:gravity_fits_additional} shows the interferometric data (colored) and best-fit binary model (black) for the remaining VLTI/GRAVITY observations. 

\section{C. Derivation of slaved precession period}
\label{app:tau_slaved_derivation}

Here we derive the slaved precession period in the case the quadrupolar moment is dominated by the rotational bulge (Eq. \ref{eq:tau_slaved}) from first principles. 

First, it is useful to rewrite the Hamiltonian due to the rotational bulge using the so-called normalized angular momentum vector $\vect{j_{\mathrm{in}}}$ with magnitude 

\begin{align}
j_{\mathrm{in}}=\sqrt{1-e_{\mathrm{in}}^2}
\end{align}

\begin{align}
&H_{\mathrm{rotate,Aa}} = - \frac{m_{Ab}k_{Aa} R_{Aa}^5 \Omega_{Aa}^2}{6 a_{\mathrm{in}}^3}\left( 3\frac{\left( \hat{\Omega}_{Aa}\cdot \vect{j_{\mathrm{in}}} \right)^2}{j_{\mathrm{in}}^5} -\frac{1}{j_{\mathrm{in}}^3}\right ) 
\end{align}

\noindent where $\hat{\Omega}_{Aa}$ is the unit vector in the direction of the spin axis of Aa. 

We also have that 

\begin{align}
\dot{\vect{j}}_{\mathrm{in}} =-\frac1{L_{A,circ}} \vect{j_{\mathrm{in}}} \times \partial_{\vect{j}_{\mathrm{in}}}{H_\mathrm{rotate,Aa}}
\end{align}

\noindent where $L_{\mathrm{A,circ}}$ is the orbital angular momentum angular momentum for the corresponding circular orbit. 

From 

\begin{align}
&\partial_{\vect{j}_{\mathrm{in}}}{H_\mathrm{rotate,Aa}} = - \frac{m_{Ab}k_{Aa} R_{Aa}^5 \Omega_{Aa}^2}{6 a_{\mathrm{in}}^3} \left(\frac{6 \left(\hat{\Omega}_{Aa}\cdot \vect{j}_{\mathrm{in}}\right)}{j_{\mathrm{in}}^5}\hat{\Omega}_{Aa}+\left(-15\frac{\left(\hat{\Omega}_{Aa}\cdot \vect{j}_{\mathrm{in}}\right)^2}{j_{\mathrm{in}}^6}+3\frac{1}{j_{\mathrm{in}}^4}\right) \hat j_{\mathrm{in}} \right)
\end{align}

it follows that 

\begin{align}
\dot{\vect{j}}_{\mathrm{in}} = \frac{m_{Ab}k_{Aa} R_{Aa}^5 \Omega_{Aa}^2}{a_{\mathrm{in}}^{3}L_{\mathrm{A,circ}}j_{\mathrm{in}}^3} \cos(i_{Aa}) (\hat j_{\mathrm{in}} \times \hat{\Omega}_{Aa})
\end{align}

The torque on Aa due to Ab is equal to that of the rotational bulge of Aa on the orbit: 

\begin{align} \label{eq:torque_spin}
\dot {\hat{\Omega}}_{Aa} =- \frac{L_{\mathrm{A,circ}} \dot{\vect{j}}_{\mathrm{in}}}{I_{Aa}\Omega_{Aa}}
\end{align}

\noindent so that 

\begin{align}
\dot {\hat{\Omega}}_{Aa} =-\frac{m_{Ab}k_{Aa} R_{Aa}^5 \Omega_{Aa}}{a_{\mathrm{in}}^{3} I_{Aa} j_{\mathrm{in}}^3}\cos(i_{Aa})(\hat j_{\mathrm{in}}\times \hat{\Omega}_{Aa})
\end{align}

So the precession rate and period are given by

\begin{align}
\Omega_{\mathrm{slaved,Aa}} = \frac{m_{Ab} k_{Aa} R_{Aa}^5  \Omega_{Aa}}{a_{\mathrm{in}}^{3} I_{Aa} j_{\mathrm{in}}^3} \cos(i_{Aa}) \\
\tau_{\mathrm{slaved,Aa}} = \frac{2\pi}{\Omega_{\mathrm{slaved,Aa}}} = 2 \pi \frac{a_{\mathrm{in}}^{3} I_{Aa} j_{\mathrm{in}}^3}{m_{Ab} k_{Aa} R_{Aa}^5  \Omega_{Aa} \cos(i_{Aa})}
\end{align}

\noindent which is the same as Eq. \ref{eq:tau_slaved} with $a_{\mathrm{in}}^{3} = G (M_{Aa}+M_{Ab}) \left ( \frac{P_{\mathrm{in}}}{2 \pi} \right )^2$, $I_{Aa} = \beta_{Aa}^2 M_{Aa} R_{Aa}^2$ and $j_{\mathrm{in}} = \sqrt{1-e_{\mathrm{in}}^2}$.

\section{D: Comparison between analytical solution and numerical solution} \label{app:numerical_solution}

Here we compare our analytical solution, which assumes that the inclinations of the stellar spin axes angles $i_{Aa}$ and $i_{Ab}$ relative to $\hat{j}_{\mathrm{in}}$ remain constant due to the fast slaved precession as detailed in section \ref{subsec:slaved_precession}, with the numerical solution obtained by integrating the equations of motion, namely 

\begin{align}
\frac{d \vect{j}}{dt} = -\frac{1}{J_{\mathrm{in,circ}}} ( \vect{e} \times \partial_{\vect{e}}H + \vect{j} \times \partial_{\vect{j}}H) \\
\frac{d \vect{e}}{dt} = -\frac{1}{J_{\mathrm{in,circ}}} ( \vect{j} \times \partial_{\vect{e}}H + \vect{e} \times \partial_{\vect{j}}H)
\end{align}

\noindent where $J_{\mathrm{in,circ}} = (G M_{\mathrm{in}} a_{\mathrm{in}})^{1/2}$, $\vect{j} = (1-e_{\mathrm{in}}^2)^{1/2} \hat{j}_{\mathrm{in}}$ and $\vect{e} = e_{\mathrm{in}} \hat{e}_{\mathrm{in}}$ \citep[e.g.][]{Tremaine09}. The evolution of the stellar spin axis follows from Eq. \ref{eq:torque_spin}. 

\begin{figure*}[]
\centering
\includegraphics[width=0.7\textwidth]{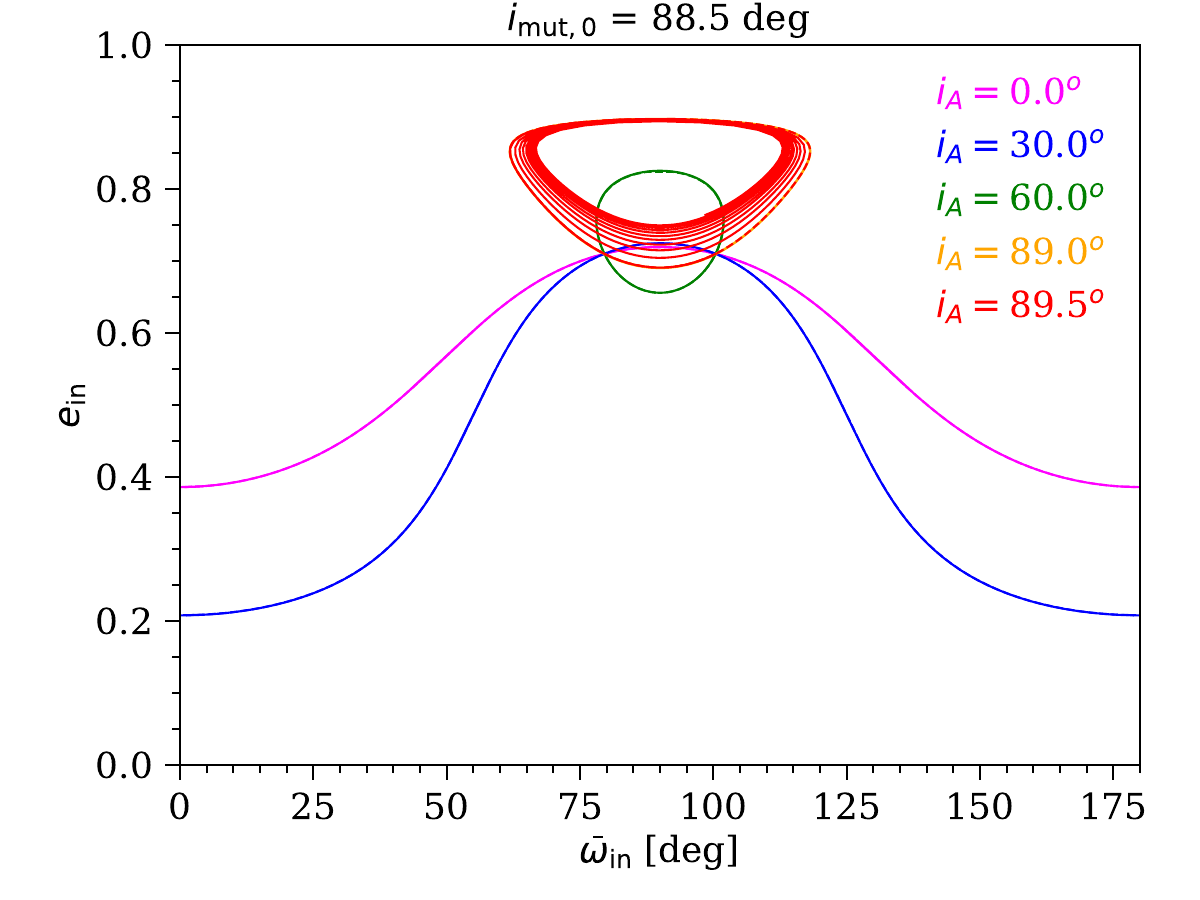} \\
\includegraphics[width=0.7\textwidth]{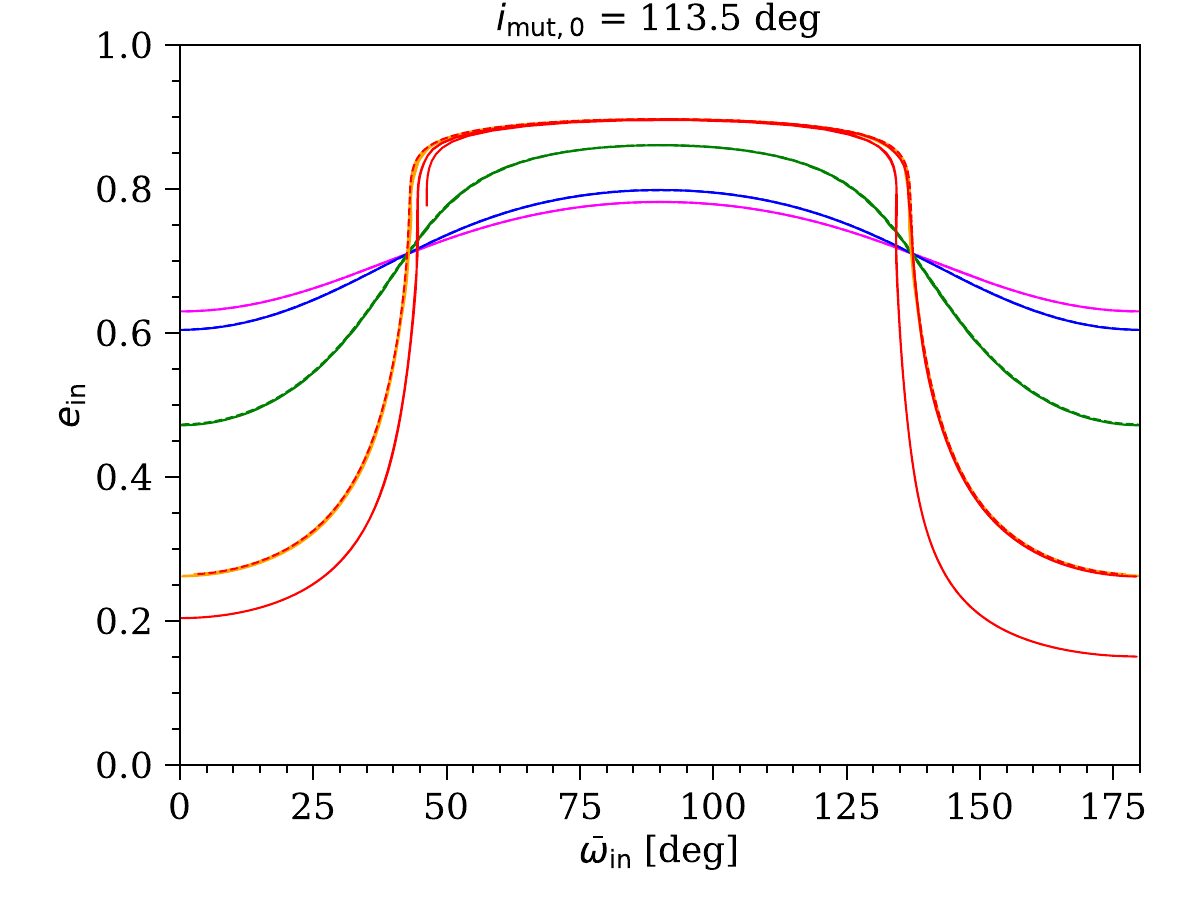}
\caption{\label{fig:analytical_vs_numerical} Dynamics plots for the inner binary in Lambda Ophiuchi for the analytical (dashed lines) and numerical (solid lines) solutions for different initial inclinations of the stellar spin axes $i_A$. The analytical and numerical solutions are indistinguishable except when $i_A \gtrsim 89^{\degr}$.}
\end{figure*}

Figure \ref{fig:analytical_vs_numerical} shows the $\bar{\omega}_{\mathrm{in}}-e_{\mathrm{in}}$ contours for the analytical and the numerical solutions for different values of $i_{Aa} = i_{Ab} = i_{A}$. In the latter case, the initial orientation of the stellar spin axes are 

\begin{align}
\hat{\Omega}_{Aa} = \cos(i_{Aa}) \hat{j}_{\mathrm{in}} + \sin(i_{A}) \cos(\phi_{Aa}) \hat{e}_{\mathrm{in}} + \sin(i_{A}) \sin(\phi_{Aa}) \hat{b}_{\mathrm{in}}
\end{align}

\noindent and similarly for $\hat{\Omega}_{Ab}$, where $\hat{b}_{\mathrm{in}} = \hat{j}_{\mathrm{in}} \times \hat{e}_{\mathrm{in}}$ so that $\phi_{Aa}$ is the orientation of the projection of $\hat{\Omega}_{Aa}$ in the orbital plane. We use $\phi_{Aa} = \phi_{Ab} = 45^{\degr}$ as this angle has no influence on the dynamics. As in Section \ref{subsec:dynamics}, we assume $v_{Aa} = 140 \text{ km}\text{ s}^{-1}$ and $v_{Ab} = 84 \text{ km}\text{ s}^{-1}$ for the rotational velocities. 

The analytical solutions are plotted with dashed lines and the numerical solutions with solid lines. For stellar spin axes inclinations $i \leq 89^{\degr}$, the two solutions are so close that they indistinguishable by eye. For $i \gtrsim 89^{\degr}$, the analytical solution with a single loop in the $e$ vs $\omega$ plane starts to deviate significantly from the numerical solution since $i_A$ is no longer approximately constant. This is fully in line with the expectations as described in Section \ref{subsec:slaved_precession}. 

\section{E: Two dimensional plots for \lowercase{$\Delta e$}} \label{app:delta_e_2d}

\begin{figure}[]
\centering
\includegraphics[width=0.5\textwidth]{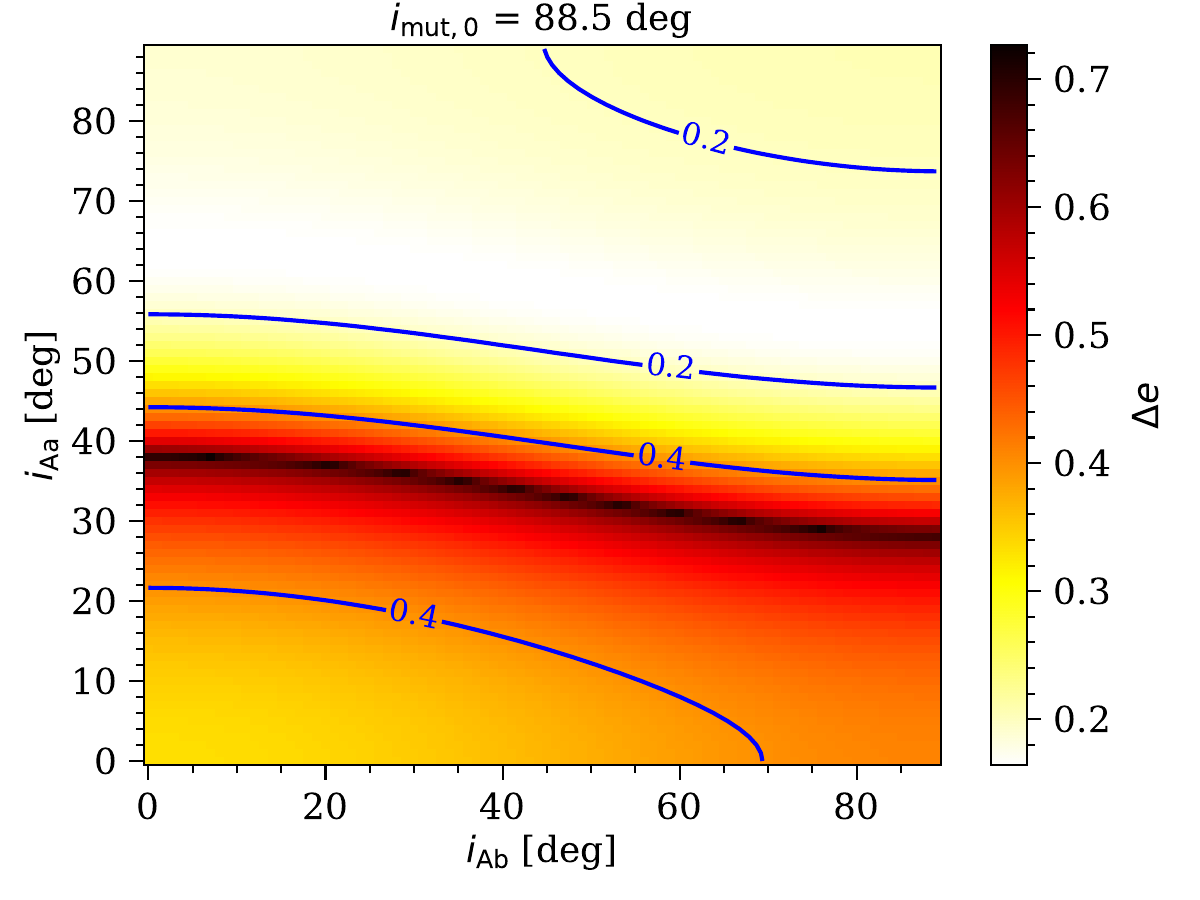} \\
\includegraphics[width=0.5\textwidth]{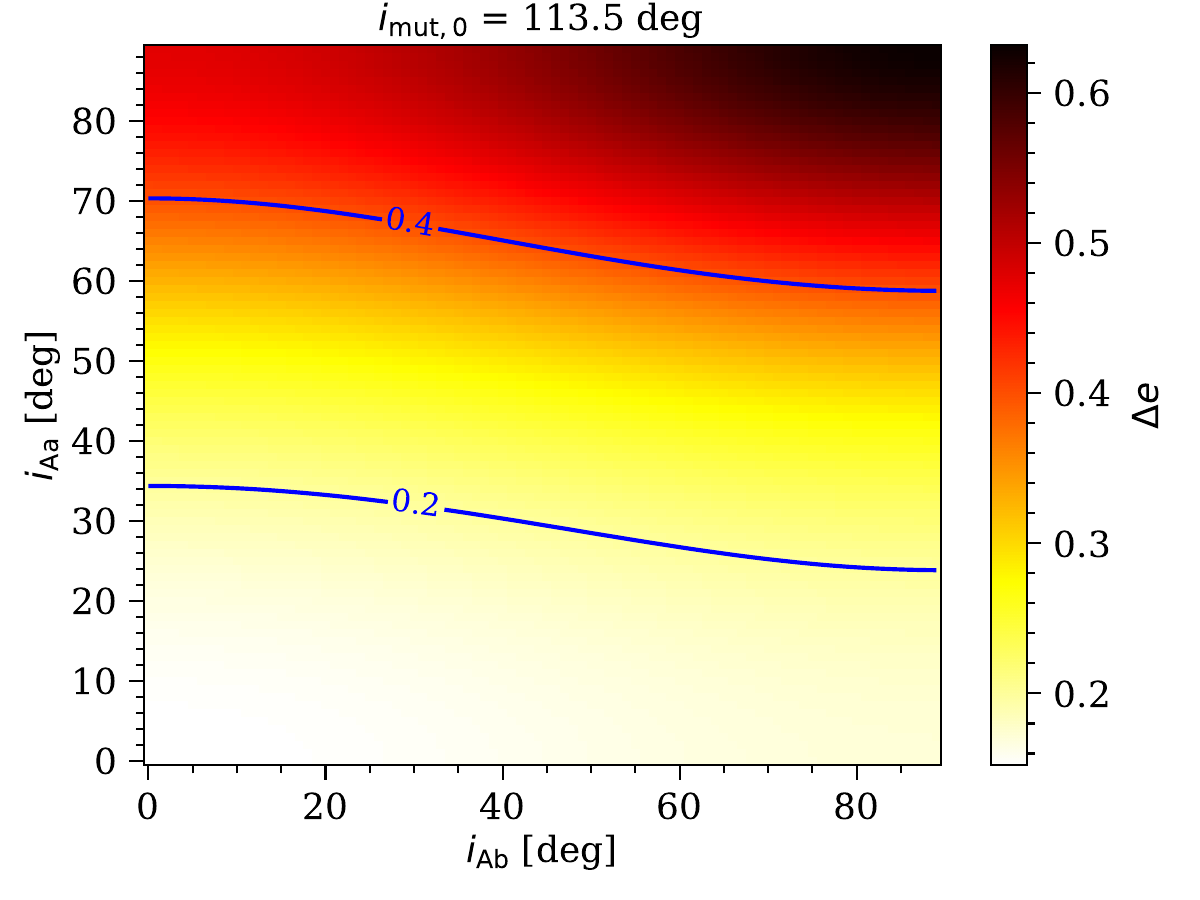}
\caption{\label{fig:delta_e_2d} Amplitude $\Delta e = e_{\mathrm{max}} - e_{\mathrm{min}}$ of the von Zeipel-Kozai oscillations in the inner binary of Lambda Ophiuchi as a function of the inclinations between the stellar spin axes and the inner orbital axis.}
\end{figure}

Figure \ref{fig:delta_e_2d} shows the amplitude of the von Zeipel-Kozai-Lidov oscillations in the inner binary of Lambda Ophiuchi as a function of both spin axes inclinations relative to the inner orbital axis. The amplitude depends mostly on $i_{Aa}$ since the rotational bulge precession is dominated by the primary due to its larger radius. 

\end{document}